%% file: main.tex
\documentclass[sigconf, nonacm]{acmart}

\newcommand\availabilityurl{https://github.com/HPI-Information-Systems/dependency-based-qo}

\usepackage[capitalise,nameinlink,noabbrev]{cleveref}

\usepackage{siunitx}
\usepackage{subcaption}
\usepackage{alltt}
\usepackage{soul}
\usepackage{datenumber}
\usepackage{xspace}

\newcommand*{\aSet}[1]{\ensuremath{\boldsymbol{\mathit{#1}}}}
\newcommand*{\aList}[1]{\ensuremath{\boldsymbol{\mathsf{#1}}}}
\newcommand*{\aTab}[1]{\textit{#1}}
\newcommand*{\aCol}[1]{\texttt{#1}}

\newcommand*{\rSize}[1]{\ensuremath{\text{size}({#1})}}
\newcommand*{\rCard}[1]{\ensuremath{\text{dist}({#1})}}

\newcommand*{\join}[4]{\ensuremath{\aTab{#2} \mathbin{#4_{#1}} \aTab{#3}}}
\newcommand*{\innerjoin}[3][]{\join{#1}{#2}{#3}{\Join}}
\newcommand*{\semijoin}[3][]{\join{#1}{#2}{#3}{\ltimes}}
\newcommand*{\rUnaryOp}[3]{\ensuremath{#3_{\aCol{#2}}(\aTab{#1})}}
\newcommand*{\selection}[2]{\rUnaryOp{#1}{#2}{\sigma}}

\newcommand*{\aggregation}[2]{\rUnaryOp{#1}{#2}{\gamma}}

\def\ojoin{\setbox0=\hbox{$\Join$}%
\rule[0.05ex]{.2em}{.4pt}\llap{\rule[1.05ex]{.2em}{.4pt}}}

\newcommand*{\leftjoin}[3][]{\join{#1}{#2}{#3}{\ojoin\mkern-6.7mu\Join}}

\newcommand*{\pEq}{\unskip\ensuremath{\,=\,}}

\newcommand*{\pIsNotNull}{\unskip\ensuremath{\,\aCol{IS}\,\aCol{NOT}\,\aCol{NULL}}}

\newcommand*{\FD}[2]{\ensuremath{#1 \to #2}}
\newcommand*{\OD}[2]{\ensuremath{#1 \mapsto #2}}
\newcommand*{\IND}[2]{\ensuremath{#1 \subseteq #2}}

\newcommand\thefont{\expandafter\string\the\font}
\newcommand*{\perc}[1]{\qty{#1}{\percent}}
\newcommand*{\circled}[1]{\raisebox{.5pt}{\textcircled{\raisebox{.3pt}{\scriptsize\textsf{#1}}}}}

\newcommand*{\mus}[1]{\qty{#1}{\unit{\micro\second}}}
\newcommand*{\ms}[1]{\qty{#1}{\unit{\milli\second}}}
\newcommand*{\s}[1]{\qty{#1}{\unit{\second}}}

\newcommand*{\lastaccess}{\setdate{2025}{04}{14}(accessed \datedate)}
\newcommand{\eg}{e.\,g.,\ }
\newcommand{\ie}{i.\,e.,\ }
\newcommand{\punctfootnote}[1]{\unskip\nolinebreak\hspace{-.2em}\footnote{#1}}
\newcommand*{\hana}{SAP HANA\xspace}
\newcommand*{\ca}[1]{\ensuremath{\approx#1}}
\newcommand*{\landO}[1]{\ensuremath{\mathcal{O}(#1)}}
\renewcommand*{\landO}[1]{\ensuremath{\text{\usefont{OMS}{cmsy}{m}{n}O}(#1)}}

\newcommand*{\myparagraph}[1]{\smallskip\noindent\textit{#1.}}

\begin{document}

\title{Enabling Data Dependency-based Query Optimization}

\author{Daniel Lindner}
\orcid{0009-0003-1849-7262}
\affiliation{%
  \institution{Hasso Plattner Institute}
  \city{Potsdam}
  \country{Germany}
}
\email{daniel.lindner@hpi.de}

\author{Daniel Ritter}
\orcid{0000-0001-6146-3365}
\affiliation{%
  \institution{SAP}
  \city{Walldorf}
  \country{Germany}
}
\email{daniel.ritter@sap.com}

\author{Felix Naumann}
\orcid{0000-0002-4483-1389}
\affiliation{%
  \institution{Hasso Plattner Institute}
  \city{Potsdam}
  \country{Germany}
}
\email{felix.naumann@hpi.de}

\begin{abstract}
Primary key (PK) and foreign key (FK) constraints are widely used for query optimization.
Knowledge about additional data dependencies, such as order dependencies, enables further substantial performance improvements.
However, such dependencies are not maintained by database systems or are even unknown to the user.
Identifying and validating relevant dependencies automatically and efficiently remains an unsolved problem.
This paper presents a system that (i)~recognizes dependency candidates for optimization, (ii)~efficiently validates their applicability to a query, and (iii)~optimizes query plans using valid dependencies.

First, we demonstrate the performance impact of optimization techniques using data dependencies beyond PKs and FKs.
Using rewritten SQL queries, we empirically show that data dependencies improve performance for a wide range of analytical database systems and benchmarks.
Second, we present how to integrate data dependencies into a system to use them without (i)~manual declaration and maintenance or (ii)~SQL rewrites.
Our integrated and fully automated system matches the performance of dedicated SQL rewrites:
compared to using only PKs and FKs, queries improve with geometric mean speedups of \perc{35} for TPC-DS and \perc{29} for JOB\@.
Individual query latencies drop by more than \perc{90}.
The dependency discovery overhead is orders of magnitude lower than the latency improvement of a single workload execution.
\end{abstract}



\maketitle

\section{Introduction}
\label{sec:intro}
Query optimization (QO) in database systems is crucial to find efficient execution plans and has been studied since the dawn of relational databases~\cite{DBLP:journals/csur/Ioannidis96,DBLP:journals/pvldb/Neumann14}.
Essential and well-known optimization techniques, such as predicate placement~\cite{DBLP:conf/sigmod/SelingerACLP79}, join ordering~\cite{DBLP:journals/tods/IbarakiK84}, and subquery unnesting~\cite{DBLP:journals/tods/Kim82}, substantially improve workload execution times.
In modern systems, more sophisticated optimization and execution strategies still improve the performance of multiple queries~\cite{DBLP:conf/icde/PerronSKS19,DBLP:journals/pvldb/JustenRFLTLBHZMB24,DBLP:journals/vldb/FentBN23}.
As part of these efforts, optimizations using data dependencies have been proposed throughout the history of database research~\cite{DBLP:journals/vldb/KossmannPN22}.

Data dependencies are
``metadata that describe relationships among
columns''~\cite[p.~561]{DBLP:journals/vldb/AbedjanGN15}, and they formalize specific properties of datasets.
Their discovery and use have been researched for decades~\cite{DBLP:conf/icalp/FaginV84} for different application areas, such as data cleaning~\cite{DBLP:journals/pvldb/RezigOAEMS21,DBLP:journals/pvldb/FanGJ08} and data integration~\cite{DBLP:conf/pods/Lenzerini02,DBLP:conf/vldb/MadhavanBR01}.
Among others, these data dependencies include unique column combinations (UCCs), functional dependencies (FDs), inclusion dependencies (INDs), and order dependencies (ODs).
For instance, a UCC states that tuples have no duplicate values for a given set of attributes, and an IND means that values for specific attributes are also present in another set of attributes, often in another table (see \cref{sec:background:dependencies} for details).
In database systems, these two dependency types can be represented using unique/primary key (PK) and foreign key (FK) constraints, and database management systems (DBMSs) often apply optimizations if these constraints are present (see \cref{sec:evaluation:systems}).

Many data dependencies are valid in real-world datasets~\cite{DBLP:conf/icse/YangSYC020,Bell95} due to implications on the modeled entities, application logic, or just by chance.
Yet, DBMSs apply only few dependency-based optimizations~\cite{DBLP:journals/vldb/KossmannPN22} based on PKs and FKs.
Previous research emphasized that valid dependencies are often unknown or not declared as constraints~\cite{DBLP:conf/ecsqaru/Bell97,DBLP:journals/pvldb/LiuWSPLJYYLC23,DBLP:conf/icse/YangSYC020}.
SQL cannot even express some relevant dependencies, such as FDs or ODs, as table constraints~\cite[p.~89]{ISO:9075:2023}.
Also, for data loaded from standardized storage formats, such as CSV, Parquet, or ORC files, no column-wide constraints can be specified, let alone constraints across multiple tables, \eg FKs/INDs.
Thus, \citet{DBLP:conf/cidr/KossmannNLP22} showcased a system to discover valid dependencies using workload information.
However, their work raises two main research questions required to close the gap to make de\-pen\-den\-cy-based QO practical:

\begin{figure}[t]
    \centering
    \includegraphics[width=\linewidth]{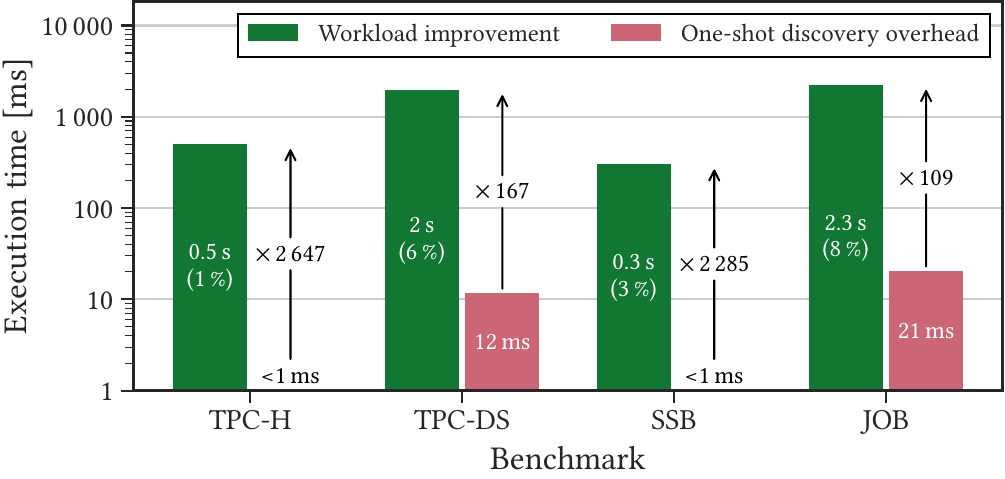}
    \caption{Workload improvement and discovery overhead when automatically discovering  and exploiting data dependencies additionally to the schema (single-threaded).}
    \label{fig:motivational2}
\end{figure}

\begin{itemize}
    \item[Q-1] How can we efficiently validate data dependencies inside database systems?
    \item[Q-2] How to thoroughly integrate data de\-pen\-den\-cy-based optimizations into a DBMS?
\end{itemize}

This paper addresses these questions by providing metadata-aware dependency validation algorithms and building blocks to integrate data dependencies as first-class citizens into a DBMS optimizer.
We perceive dependencies as sole metadata and abolish the necessity to model, enforce, and maintain constraints explicitly (\eg by creating indexes).
We select three cost-in\-de\-pen\-dent dependency-based query rewrites for groupings and joins~\cite{DBLP:journals/vldb/KossmannPN22}, which are expensive operations~\cite{DBLP:journals/pvldb/DreselerBRU20}, and apply them to different workloads and DBMSs.
Two of these rewrites rely on PKs and FKs, but we show that optimization techniques using further dependencies notably benefit performance.

Using our workload-driven architecture (\cref{sec:approach}), we discover relevant dependencies with validation algorithms tailored to databases (Q-1) as part of an \emph{extract, transform, load} (ETL) pipeline.
\Cref{fig:motivational2} shows that implementing optimizations using additional dependencies in a DBMS optimizer further improves the execution time of two benchmark workloads:
17~TPC-DS and 66~JOB queries improve with \perc{35} and \perc{29} geometric mean speedups.
Simultaneously, the discovery overhead is orders of magnitude smaller than the latency improvement of a single workload execution.

While we evaluate three concrete optimization techniques, we present an extensible framework for applying further optimizations based on UCCs, INDs, and ODs.
Additional optimization techniques can be easily added. 
We integrate the rewrites into the optimizer of an open-source DBMS as cost-independent transformations.
As a precondition, we propagate dependencies in the query plan and modify the optimizer and execution engine to handle subqueries, transforming joins into selections (Q-2).

After discussing related work in \cref{sec:related-work}, \cref{sec:background} introduces fundamental concepts of data dependencies, dependency-based query optimization, and database systems.
\Cref{sec:approach} presents our architecture.
In particular, we make the following contributions:
\begin{itemize}
    \item[C-1] \emph{Metadata-aware dependency validation}. We propose algorithms that exploit data layout, encoding, and statistics to achieve negligible overhead when validating four dependency types (\cref{sec:validation}).
    \item[C-2] \emph{Dependency propagation}. We describe how to represent and propagate dependencies in the query plan to enable the application of dependency-based QO techniques (\cref{sec:propagation}).
    \item[C-3] \emph{Subquery handling}.
    One rewrite introduces scalar subqueries in selections.
    Thus, we handle these subqueries 
    in the DBMS, focusing on cardinality estimation and dynamic pruning using subquery results during execution (\cref{sec:subqueries}).
    \item[C-4] \emph{Evaluation}. We evaluate the potential of dependency-based optimization 
    even beyond using PKs and FKs for four state-of-the-art systems.
    Furthermore, we analyze the benefits of system integration and discuss the overhead of additional dependency discovery (\cref{sec:evaluation}).
    Finally, we discuss which workloads benefit most.
\end{itemize}

\noindent We conclude and discuss how future work can extend our approach to frequently changing datasets in \cref{sec:conclusion}.
Our open-source implementation is available online.\punctfootnote{See \url{\availabilityurl}.}

\section{Related Work}
\label{sec:related-work}
We identify two main research fields connected to metadata-based query optimization.
First, data dependency-based query optimization has been proposed for various systems and recently applied to a research prototype.
Second, data profiling systems for the automatic discovery of further metadata have been developed to use discovered metadata for query optimization.

\subsection{Data Dependency-based Query Optimization}
\label{sec:related-work:dependencies}
Query optimization using data dependencies has been proposed since the 1970s.
In their survey, \citet{DBLP:journals/vldb/KossmannPN22} collected more than 60 such optimization techniques, grouping them by the type of exploited dependency, the affected operator of the relational algebra, and the optimization category.
\Cref{sec:background:rewrites} presents three logical query rewrites~\cite{DBLP:conf/tpctc/BonczNE13,DBLP:books/aw/DateD92,mySQLsemiJoin,DBLP:conf/sigmod/AbadiMH08,DBLP:conf/edbt/SzlichtaGGMPZ11} in detail.

Based on their survey, \citet{DBLP:conf/cidr/KossmannNLP22} presented an approach to automatically generate and validate interesting dependency candidates using workload information.
They evaluated to which extent selected optimizations using these discovered dependencies benefit the performance of three benchmark workloads.
An overview of their system is given in \cref{sec:approach}.
However, that work provided only a proof of concept, where dependency-based optimizations were not fully integrated into the DBMS\@.
Furthermore, the validation algorithms caused considerable overhead compared to the performance improvement per workload execution.
Thus, the work left two main challenges unresolved:
(i)~actual integration of dependency-based QO into the system and
(ii)~efficient validation of dependency candidates.

We address these issues by thoroughly integrating selected optimization techniques in the query processing pipeline, extending query optimization and execution logic.
Thus, we show that system integration improves performance characteristics (\cref{sec:evaluation:performance}).
Furthermore, we present novel, highly optimized metadata-aware dependency validation strategies and evaluate the impact of selected optimization techniques for different DBMSs.

\subsection{Further Metadata for Query Optimization}
\label{sec:related-work:metadata}
Data profiling refers to the task of metadata discovery~\cite{DBLP:journals/vldb/AbedjanGN15}.
Traditional data dependency mining algorithms aim to find all valid dependencies in a given dataset.
Efficient algorithms have been proposed for different dependency types in single-node and distributed environments~\cite{DBLP:journals/pvldb/BirnickBFNPS20,DBLP:journals/pvldb/PapenbrockEMNRZ15,DBLP:conf/cikm/DurschSWFFSBHJP19}.
However, (i)~finding all dependencies in a dataset is expensive, and (ii)~further metadata besides data dependencies can be used for query optimization.

Thus, various systems exist to discover and use \emph{semantic constraints} for cost-based query optimization~\cite{DBLP:conf/sigmod/King80}.
An example of such a semantic constraint is that every manager in a company is paid a bonus of at least \$\num{1000}.
\citet{DBLP:journals/tkde/YuS89} and \citet{DBLP:books/mit/fayyadPSU96/HsuK96} compared the result sets of queries to derive valid constraints.
Thus, they could only transform queries if relevant query reformulations were also part of the workload.
To overcome this shortcoming, \citet{DBLP:journals/tkde/ShekharHKC93} and \citet{DBLP:journals/jidm/PenaFMA18} derived valid constraints from the data first and used them for optimization in the second step.
These systems also discover semantic constraints that cannot be used for query optimizations, leading to avoidable overhead.
Furthermore, they add another optimization layer on top of the DBMS\@.
\citet{DBLP:journals/tods/SiegelSS92} generated semantic constraint candidates during optimization and validated them later, coupling constraint discovery tightly with query optimization.

Recently, \citet{DBLP:journals/pvldb/LiuWSPLJYYLC23} performed static source code analysis to identify various constraints the applications guarantee, such as inclusion dependencies, regular expressions for strings, or attribute nullability.
They used these constraints for SQL query preprocessing, stating that ``most of the inferred constraints [were] not declared in the database,'' optimizers did not support specific rewrites, and they were ``unaware of any existing tools that can discover [constraints]''~\cite[p.~1209--1210]{DBLP:journals/pvldb/LiuWSPLJYYLC23}.

In contrast, we present a system to integrate dependency-based optimization techniques and the discovery of beneficial dependencies into database systems.
Instead of semantic constraints, we exploit data dependencies.
Furthermore, our approach to collecting relevant metadata is decoupled from the core query execution and uses specialized  validation algorithms.

\section{Data Dependencies for Optimization}
\label{sec:background}
This section describes the basic concepts we build upon in our work.
After we define different data dependency types in \cref{sec:background:dependencies}, \cref{sec:background:rewrites} illustrates three dependency-based logical query rewrites using an example query.
Finally, \cref{sec:background:databases} introduces relevant features of relational DBMSs.

\subsection{Data Dependencies}
\label{sec:background:dependencies}
Data dependencies are dedicated metadata that describe how data is interrelated.
Specific relationships are formalized to prove and compute whether a dataset fulfills a dependency's requirements, \ie whether the dependency is \emph{valid} or not.
In the following, we define four types of data dependencies.

\myparagraph{Unique column combination (UCC)}
Let \aTab{R} be a relation. The subset of attributes $\aSet{X} \subseteq \aTab{R}$ is a UCC iff there are no tuples whose projection on \aSet{X} is equal~\cite{DBLP:journals/jcss/LucchesiO78}.
UCCs can occur by chance or stem from real-world identifiers or surrogate keys.
Thus, they are also referred to as \emph{candidate keys}~\cite{DBLP:persons/Codd71b}.
Relational databases can enforce UCCs via unique or primary key constraints.

\myparagraph{Functional dependency (FD)}
An FD \FD{\aSet{X}}{\aSet{Y}} is valid iff all tuples with the same values for $\aSet{X} \subseteq \aTab{R}$ also have the same values for $\aSet{Y} \subseteq \aTab{R}$~\cite{DBLP:persons/Codd71b,DBLP:books/cs/Ullman88}.
In particular, the FD \FD{\textit{X}}{\aTab{R} \setminus \aSet{X}} always holds if \aSet{X} is a UCC\@.
Real-world relationships often cause FDs, \eg \FD{\aCol{zip}}{\aCol{city}}.

\myparagraph{Order dependency (OD)}
If ordering the tuples of \aTab{R} by \aList{X} also orders the tuples by \aList{Y}, then \OD{\aList{X}}{\aList{Y}} is a valid OD~\cite{DBLP:journals/pvldb/SzlichtaGG12}.
In this case, \aList{X} and \aList{Y} are lists of attributes in \aTab{R}, \ie the attribute order is relevant.
ODs often occur in data that includes a time component~\cite{DBLP:conf/edbt/SzlichtaGGMPZ11,DBLP:journals/pvldb/SzlichtaGG12}.

\myparagraph{Inclusion dependency (IND)}
The IND \IND{\aList{X}}{\aList{Y}} is valid iff all distinct values of $\aTab{R}[\aList{X}]$ are also present in $\aTab{S}[\aList{Y}]$~\cite{DBLP:conf/vldb/CasanovaTF88}. As a special case, \aTab{R} and \aTab{S} might refer to the same relation.
INDs often represent membership or ownership and can be enforced by a foreign key constraint.

\subsection{Data Dependency-based Query Rewrites}
\label{sec:background:rewrites}
Corresponding to the work of \citet{DBLP:conf/cidr/KossmannNLP22},
we picked a subset of three dependency-based logical query rewrites in their survey~\cite{DBLP:journals/vldb/KossmannPN22}  that rely on the four dependency types defined in \cref{sec:background:dependencies}.
Specific query rewrites promise to improve performance always~\cite{DBLP:journals/vldb/KossmannPN22}.
Furthermore, the selected rewrites target aggregate and join operators, which are
costly for analytical workloads~\cite{DBLP:journals/pvldb/DreselerBRU20}.
To illustrate the three rewrites, we use an example query inspired by TPC-DS data and constraints, which selects each customer's ID, name, and the sum spent on purchases for a specific time period:
\begin{alltt}
\hfill\hphantom{  }\textbf{SELECT} c_sk, c_name, sum(s_sales_price)\hphantom{        }\hfill
\hfill\hphantom{    }\textbf{FROM} date_dim\hphantom{                                }\hfill
\hfill\hphantom{    FROM} \textbf{INNER JOIN} sales \textbf{ON} d_sk = s_sold_date\hphantom{  }\hfill
\hfill\hphantom{    FROM} \textbf{INNER JOIN} customer \textbf{ON} s_customer = c_sk\hfill
\hfill\hphantom{   }\textbf{WHERE} d_date = '2000-01-01'\hphantom{                   }\hfill
\hfill\textbf{GROUP BY} c_sk, c_name;\hphantom{                           }\hfill
\end{alltt}

\begin{figure*}[t]
    \centering
    \begin{subfigure}{.33\linewidth}
        \centering
        \includegraphics[height=1.51in]{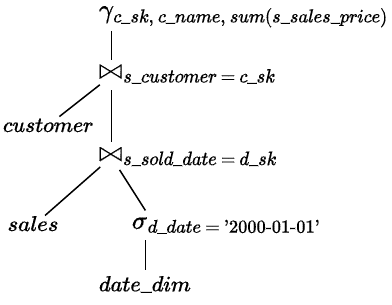}
        \caption{Original plan}
        \label{fig:rewrites-subfig:base}
    \end{subfigure}\hfill%
    \begin{subfigure}{.33\linewidth}
        \centering
        \includegraphics[height=1.51in]{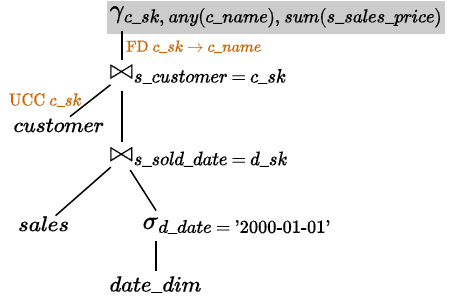}
        \caption{Dependent group-by reduction (O-1)}
        \label{fig:rewrites-subfig:o1}
    \end{subfigure}\hfill%
    \begin{subfigure}{.33\linewidth}
        \centering
        \includegraphics[height=1.51in]{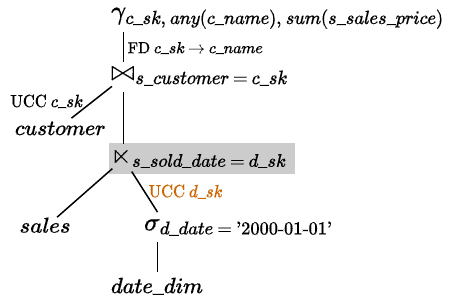}
        \caption{Join-to-semi-join rewrite (O-2)}
        \label{fig:rewrites-subfig:o2}
    \end{subfigure}
    \\[2ex]
    \begin{subfigure}{.5\linewidth}
        \centering
        \includegraphics[height=2in]{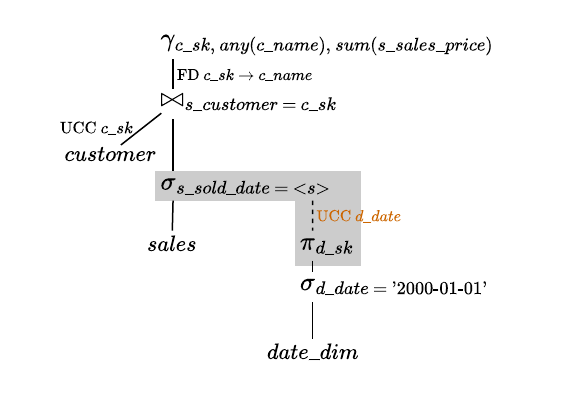}
        \caption{Join-to-predicate rewrite (i)~(O-3)}
        \label{fig:rewrites-subfig:o3i}
    \end{subfigure}\hfill%
    \begin{subfigure}{.5\linewidth}
        \centering
        \includegraphics[height=2in]{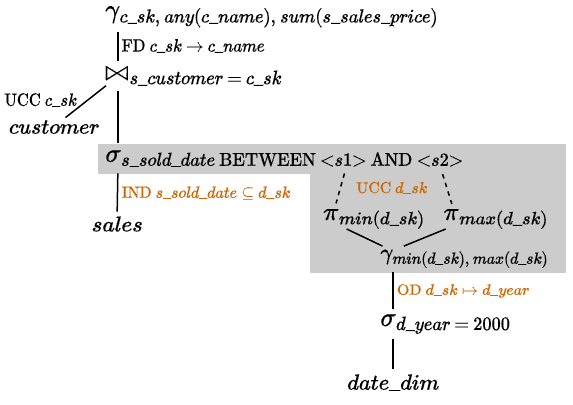}
        \caption{Join-to-predicate rewrite (ii)~(O-3)}
        \label{fig:rewrites-subfig:o3ii}
    \end{subfigure}
    \caption{Original query plan and versions successively rewritten using O-1, O-2, and O-3. Edges are annotated with the data dependencies that enable the following rewrites. Rewritten parts and dependencies used are highlighted. Note that the selection predicate on \emph{date\_dim} was changed to showcase the OD-based version of O-3.}
    \label{fig:rewrites}
\end{figure*}

\Cref{fig:rewrites} shows query plans resulting from the three query rewrites, where \cref{fig:rewrites-subfig:base} is the original query plan.
Dependencies exploited by individual optimization techniques are highlighted.

\myparagraph{O-1 Dependent group-by reduction~\cite{DBLP:conf/tpctc/BonczNE13,DBLP:books/aw/DateD92}}
Grouping by multiple attributes can be avoided if an FD's determinant \emph{and} dependent attributes are part of the group-by list.
We remove all dependent attributes from the grouping set and select any dependent attribute value of dependent attributes, as they are uniform within the group.
In the example query, the customer's name \texttt{c\_name} is unique for their ID \texttt{c\_sk}.
Thus, we only group by \texttt{c\_sk}.

\myparagraph{O-2 Join-to-semi-join rewrite~\cite{mySQLsemiJoin}}
The second query rewrite transforms an inner equi-join \innerjoin{R}{S} to a semi-join \semijoin{R}{S}.
Many DBMSs implement semi-joins as they execute them efficiently
~\cite{DBLP:conf/vldb/AhmedLWDSZC06,mySQLsemiJoin,DBLP:conf/vldb/GraefeBC98,DBLP:conf/sigmod/BandleG021,DBLP:conf/btw/May0L17}.
This rewrite is possible if \aTab{S}'s join key is unique and subsequent operators or the final projection require no further attributes of \aTab{S}.
In fact, the semi-join acts as a filter for \aTab{R} by the values of \aTab{S}'s join key(s).
\Cref{fig:rewrites-subfig:o2} shows that we perform a semi-join to replace \innerjoin{sales}{date\_dim}. Here,
\aTab{date\_dim}'s join key is the primary key, and no attribute of \aTab{date\_dim} is selected later.

\myparagraph{O-3 Join-to-predicate rewrite~\cite{DBLP:conf/sigmod/AbadiMH08,DBLP:conf/edbt/SzlichtaGGMPZ11}}
If joins are merely used to filter relations, we might even replace them with a selection.
In our example query, the \aTab{date\_dim} table represents each day.
Thus, the \aCol{d\_date} column is unique.
Selecting a single day results in a single value for the join key.
Instead of joining the \aTab{sales} table with the \aTab{date\_dim} table, the query plan of \cref{fig:rewrites-subfig:o3i} turns the join into a selection to filter \aTab{sales} for the single join key, which is determined by a scalar subquery.

Similarly, the rewrite can be applied to range predicates.
Adapting our example query, we change the temporal filter from \texttt{d\_date = '2000-01-01'} to \texttt{d\_year = 2000}.
The OD \OD{\aCol{d\_sk}}{\aCol{d\_date}} ensures that the minimal and maximal join keys within the selected \aCol{d\_date} values are fed into the join, and the combination of the  IND \IND{\aCol{s\_sold\_date}}{\aCol{d\_sk}} and the UCC \aCol{d\_sk} guarantees that all tuples of the \aTab{sales} table have exactly one join partner.
Thus, we can rewrite the join to a selection with a predicate value between the minimum and maximum of the join key in \cref{fig:rewrites-subfig:o3ii}.

\subsection{Relevant Database Concepts}
\label{sec:background:databases}
Dependency-based query optimization is applicable to any DBMS\@.
However, columnar, partitioned, and encoded data with statistics is the basis of our novel dependency validation algorithms and  dynamic partition pruning.
We integrate the dependency discovery system as a plug-in as an optional task decoupled from the DBMS core.
Thus, we explain these pertinent DBMS concepts in the following paragraphs.

\myparagraph{Storage layout}
Many commercial, open-source, and research DBMSs support columnar storage~\cite{DBLP:conf/sigmod/RaasveldtM19,DBLP:journals/pvldb/RamanABCKKLLLLMMPSSSSZ13,DBLP:journals/pvldb/LarsonBHHNP15,DBLP:journals/debu/IdreosGNMMK12,DBLP:journals/sigmod/FarberCPBSL11,DBLP:journals/pvldb/LiuWSPLJYYLC23,DBLP:journals/pvldb/MukherjeeCCDGHH15,DBLP:conf/edbt/DreselerK0KUP19}
to improve performance for analytical workloads~\cite{DBLP:conf/sigmod/AbadiMH08,DBLP:conf/sigmod/BinnigHF09}.
Standardized storage formats, such as Apache Parquet and ORC, also build upon this layout~\cite{DBLP:journals/pvldb/ZengHSPMZ23}.
Columns are usually split into horizontal \emph{partitions} (also called \emph{chunks} or \emph{row groups)} to ease parallelization and the distribution of large data.
Each partition contains one \emph{segment} for each column in the table, storing a fraction of the attribute's fields.

Immutable segments can be encoded to improve space and execution efficiency using light- or heavyweight compression schemes.
Dictionary encoding is often the default for real-world data~\cite{DBLP:journals/pvldb/ZengHSPMZ23,DBLP:journals/debu/FarberMLGMRD12}:
the (often sorted) dictionary stores all unique values
locally for each segment or globally for the entire column~\cite{DBLP:conf/sigmod/BinnigHF09}, and
the attribute vector references the dictionary offset for each segment position's value.

\myparagraph{Statistics}
Databases use segment statistics to refine access to stored data.
Segments' minimal and maximal values (\emph{zone maps}~\cite{DBLP:journals/pvldb/ZiauddinWKLPK17}) or value ranges (\emph{range sets}~\cite{DBLP:journals/pvldb/OrrKC19}) enable \emph{partition pruning}, \ie skipping partitions if they cannot match selection predicates.
Pruning is effective if data is partitioned by attributes that are frequently filtered, where tuples within the same value range are stored in the same partition.
Statistics are also available for Parquet files~\cite{DBLP:journals/pvldb/ZengHSPMZ23}.

\myparagraph{Plug-in interface}
Plug-in interfaces allow functionality to be added without changing the core database code~\cite{DBLP:conf/edbt/DreselerK0KUP19,DBLP:conf/cidr/Atwal24}.
Plug-ins are shared libraries that can be dynamically loaded and unloaded.

\section{Workload-driven Data Dependency Discovery}
\label{sec:approach}
This section describes the general approach of workload-driven dependency discovery and details of our implementation for an open-source DBMS.
The architecture of the workload-driven dependency discovery system is inspired by \citet{DBLP:conf/cidr/KossmannNLP22}, but replaces the main components.

\subsection{Overview}
\label{sec:approach:overview}
\begin{figure}[tb]
    \centering
    \def\svgwidth{\linewidth}
    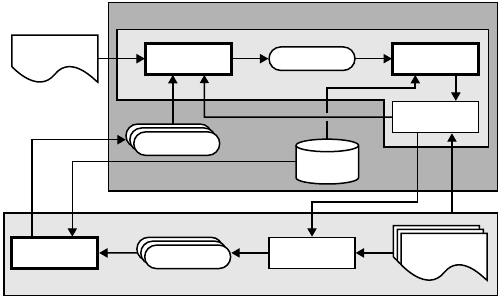
    \caption{Architectural overview of the automatic dependency discovery, based on \citet{DBLP:conf/cidr/KossmannNLP22}. We contribute to components with bold outlines.}
    \label{fig:architecture}
\end{figure}

\Cref{fig:architecture} gives an architectural overview of our automatic dependency discovery system.
During regular workload execution, the DBMS translates a SQL query into a query plan and optimizes it \circled{1}.
Optimizer rules can use metadata, such as data dependencies \circled{2}, in the optimization step.
If the same query has been issued before, the query plan is obtained from the plan cache \circled{3}.

The automatic dependency discovery is triggered during an ETL process.
Our discovery system obtains the workload's collected query plans from the plan cache \circled{4}.
The candidate generator parses these plans \circled{5} and obtains a set of dependency candidates \circled{6}.
These dependency candidates are determined by an extensible set of candidate rules \circled{7}, which anticipate the dependency-based optimizer rules' behavior and return those dependencies from which optimization could benefit if they were valid.

Candidates are validated \circled{8} on the stored data instance, skipping already validated candidates.
We describe the tailored dependency validation algorithms and the order of candidate validation in \cref{sec:validation}.
Valid dependencies are persisted \circled{9} as table metadata.
After execution, the plug-in invalidates the SQL plan cache entries of affected queries \circled{10}.
Thus, future queries are optimized again, this time with optimization techniques using the persisted dependencies~\circled{2}.
Each new optimization rule requires only a new candidate generation rule to seamlessly integrate into the system.

\subsection{Implementation}
\label{sec:approach:integration}
We implemented selected optimization techniques and dependency discovery for a modifiable DBMS with relevant features from \cref{sec:background:databases}, allowing reproducibility.
We chose the open-source system \emph{Hyrise}~\cite{DBLP:conf/edbt/DreselerK0KUP19} because it is an analytical system designed for processing high-load workloads and serving multiple concurrent users, which is a common requirement for interactive applications in the real world~\cite{DBLP:conf/icde/SethiTSPXSYJHSB19,DBLP:journals/pvldb/RenenHPVDNLSKK24,DBLP:conf/sigmod/ArmenatzoglouBB22}.
Hyrise is a columnar in-memory DBMS with horizontal partitioning into fixed-sized chunks of \num{65535} tuples,
featuring different encodings (the default is dictionary encoding), zone maps or range sets per column segment, a plug-in interface, and a rule-based optimizer with heuristic and cost-based transformations.

We contribute multiple components based on the architecture of \citet{DBLP:conf/cidr/KossmannNLP22}.
First, we facilitate using discovered data dependencies in the core DBMS\@.
\Cref{sec:propagation} explains how we propagate data dependencies in query plans to derive valid dependencies for each operator in the optimization phase.
Second, \cref{sec:subqueries} provides the adaptations required to support subqueries introduced by dependency-based rewrites.
These adaptations include improved cardinality estimation and extensions to query scheduling and execution to enable dynamic partition pruning.
Third, we present our highly optimized dependency validation algorithms in \cref{sec:validation}.

The dependency discovery's design as a plug-in decouples it from the DBMS core, making it completely optional if all dependencies are known in advance.
Triggering the discovery process is controllable and asynchronous, avoiding overhead in the running system.
Thus, dependency discovery can be performed as a one-shot overhead for static datasets, regularly as part of an ETL process, or continuously for evolving workloads, where different query templates are queried over time.
We perceive (discovered) data dependencies as additional metadata useful for optimization rather than as (SQL) data constraints.
Because dependencies, as opposed to constraints, are not enforced by the DBMS, we avoid the overhead of, \eg additional index structures and data checks regarding memory consumption and insertion latency.

\section{Metadata-aware Data Dependency Validation}
\label{sec:validation}
Reducing the overhead of additional dependency discovery requires efficient dependency validation strategies.
This section presents tailored dependency validation algorithms we designate as \emph{metadata-aware validation} (C-1).
First, we motivate the need for fast dependency validation for query optimization and explain how it differs from traditional data profiling.
Subsequently, we provide details on how we tailored algorithms to validate four types of data dependencies specifically inside a database. 
Finally, we explain how we order dependency candidates to minimize validation overhead.

We can validate specific dependency types using SQL~\cite{Bell95,DBLP:journals/vldb/AbedjanGN15}.
For instance, the following query validates the UCC candidate \aTab{R}.\aCol{a}: 
\begin{alltt}
\hfill\textbf{SELECT} count(\textbf{DISTINCT} a) = count(a) \textbf{FROM} R;\hfill
\end{alltt}

However, specialized validation algorithms outperform such validation using general-purpose database operators~\cite{DBLP:journals/vldb/AbedjanGN15,DBLP:conf/cikm/DurschSWFFSBHJP19}.
Contrary to state-of-the-art data profiling algorithms, we do not need to discover and validate \emph{all} dependencies of a particular type, which is an NP-hard problem~\cite{DBLP:journals/vldb/AbedjanGN15}.
Furthermore, we can exploit metadata and encoding characteristics provided by the database system.

Rather than optimizing traversing the search space of possible dependencies (lattice) by aggressive pruning and using data structures to combine already computed results, we focus on the efficient validation of individual dependency candidates.
We implemented our approach for an in-memory database system, mainly relying on dictionary encoding.
However, our algorithms apply to any system that uses common, accurate statistics for columns or horizontal partitions of columns, including standardized storage formats, \eg Apache Parquet. 
As the applied query rewrites mostly target joins, our tailored validation algorithms provide specializations for numeric key candidates.

In the following subsections,
\aTab{R} denotes a relation, \aCol{a} an attribute of \aTab{R}, and $\mathcal{S}_\aCol{a}$ the set of \aCol{a}'s segments, \ie partitions of \aCol{a}.
We denote the minimum and maximum attribute value present in a segment $s \in \mathcal{S}_a$ with $\min(s)$ and $\max(s)$.
The cardinality (number of distinct values) of $s$ is \rCard{s}, whereas the number of tuples in $s$ is \rSize{s}.
The notions of cardinality and size also apply to attributes and relations.
$\vert \mathcal{S}_\aCol{a} \vert$ is the number of column \aCol{a}'s segments.

\subsection{Unique Column Combinations}
\label{sec:validation:ucc}
State-of-the-art UCC discovery algorithms intersect so-called position list indexes (PLIs)~\cite{DBLP:journals/cj/HuhtalaKPT99} to combine the values of multiple columns and traverse the lattice efficiently~\cite{DBLP:journals/pvldb/BirnickBFNPS20,DBLP:conf/btw/PapenbrockN17}.
However, we can simplify the validation to construct a hash set containing a column's values for a single unary UCC candidate.
As soon as we add a value to this set without increasing the set size, the column is not unique, and we can invalidate the candidate.
If we added all fields of a column without aborting, the column is a UCC\@.

We can further optimize the validation logic by incorporating metadata known by the database.
We use a segment's minimal and maximal value, size, and cardinality to reduce validation overhead.
By accessing the first and the last elements of segments $s$'s local dictionary, we can obtain $\min(s)$ and $\max(s)$.
The dictionary size equals the number of the segment's distinct values \rCard{s}, whereas the length of the attribute vector is the number of tuples \rSize{s}.
If the segment is not dictionary-encoded or the dictionary is not sorted, zone maps, range sets, or other statistics provide this information.
For instance, Apache Parquet files also contain segment size and cardinality.\punctfootnote{See \url{https://parquet.apache.org/docs/file-format/metadata/} \lastaccess.}

\begin{figure}[tb]
    \centering
    \def\svgwidth{\linewidth}
    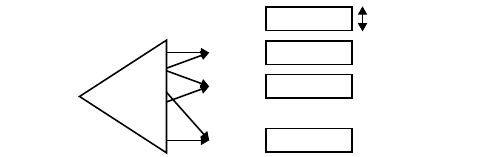
    \caption{Metadata-aware UCC validation using the on-the-fly segment index. Segment~17 invalidates the UCC: it is not unique and its domain overlaps with Segment 2.}
    \label{fig:validation_approach}
\end{figure}

For metadata-aware UCC validation, we iterate over the dictionaries or data statistics.
\Cref{fig:validation_approach} shows a running example of our UCC validation approach.
If a single segment is not unique, neither is the column.
Thus, we compare the distinct value count of each segment with its size, \ie the number of stored tuples.
We can immediately terminate the validation and reject the candidate if $\rCard{s} \neq \rSize{s}$.
This is the case for Segment~17: it has six unique values, where the segment size is seven. 
    
Furthermore, a column is unique if all segments are unique and the segments' value domains do not overlap. 
Thus, we build a segment index containing chunk IDs on the fly.
We continuously insert each segment's chunk ID with both $\min(s)$ and $\max(s)$ as keys.
We use a tree-based index for this index to iterate the nodes in a sorted manner, which
allows accessing the entries in a sorted fashion.
For each chunk ID at key $\min(s)$, the chunk ID at the following key must reference the same segment, \ie the same ID\@.
Segments~1 and~2 in \cref{fig:validation_approach} have no overlapping domains, but the minimal value of Segment~17 is between Segment~2's minimal and maximal values.
Thus, the chunk ID at key~8 differs from that at the subsequent key~12. 

However, the column can still be unique if all segments are unique, but their domains overlap.
Then, we must fall back to constructing the hash set of all values.
Segments of range-partitioned (especially sorted) primary keys do not overlap by design, speeding up their validation (cf.~\cref{sec:evaluation:validation}).
For such a range-partitioned column \aCol{a} of relation \aTab{R}, checking each segment's uniqueness and building the index has a complexity in \landO{\vert \mathcal{S}_\aCol{a} \vert \cdot \log  \vert \mathcal{S}_\aCol{a} \vert}\footnote{We assume an amortized complexity of \landO{1} for accessing the next element in the index. In particular, incrementing the tree's iterator is constant if \aCol{a} is sorted, as we always insert keys at the tree's leaf with the highest value.} rather than \landO{\rSize{\aTab{R}}} for hash set construction,\punctfootnote{For simplicity, we assume average hash set insertion in \landO{1}, \ie no hash collisions or hash table resizing. We pre-allocate the hash table to guarantee enough buckets.} where $\vert \mathcal{S}_\aCol{a} \vert  \ll  \rSize{\aTab{R}}$.

\subsection{Functional Dependencies}
\label{sec:validation:fd}
Our approach to validate FD candidates uses a simplified strategy exploiting that \FD{\aCol{a}}{\aTab{R}\setminus\aCol{a}} is a valid FD if \aCol{a} is a UCC, \eg if \aCol{a} is a primary key.
Instead of searching for FDs in all combinations of the candidate columns (\ie the \emph{lattice}), we only check if one of the columns is unique.
This simplification comes with the downside of falsely rejecting valid n-ary FD candidates with more than one determinant column.
Indeed, we miss query optimization opportunities with these candidates, but we avoid the expensive lattice traversal.
However, the anticipated query rewrite benefits most when we can reduce to a single grouping attribute.

\subsection{Order Dependencies}
\label{sec:validation:od}
We cannot avoid sorting when validating an OD candidate \OD{\aCol{a}}{\aCol{b}}.
The basic approach is to sort by \aCol{a} using the DBMS's sort operator and to verify whether \aCol{b} is also sorted.
If this is not the case, we reject the candidate.
To optimize the validation and reject invalid ODs faster, we first sort and check on a small sample.
A sample size of 100~tuples is sufficient to reject all invalid ODs in our benchmark data (\cref{sec:evaluation:validation}).

For tables with multiple chunks/partitions, we construct one segment index for each \aCol{a} and \aCol{b}.
If we iterate both indexes simultaneously and the chunk IDs have the same order, we can sort each chunk individually and only fall back to sorting the entire column if there are overlaps.
For segments of \aCol{b}, overlaps of one value, \ie $\max(s_i) = \min(s_j)$, are allowed.
In this way, we can reduce the complexity from \landO{\rSize{\aTab{R}} \cdot \log \rSize{\aTab{R}}} to $\landO{\vert \mathcal{S}_\aCol{a} \vert \cdot \log \vert \mathcal{S}_\aCol{a} \vert + \vert \mathcal{S}_\aCol{a} \vert \cdot c \cdot \log c}  \approx \landO{\vert \mathcal{S}_\aCol{a} \vert \cdot \log \vert \mathcal{S}_\aCol{a} \vert + \rSize{\aTab{R}} \cdot \log c}$, where $c$ is the chunk size (\eg fixed size of \num{65535} for Hyrise) and $\vert \mathcal{S}_\aCol{a} \vert \ll \rSize{\aTab{R}}$.
Sorting can be omitted if the partitions are already sorted by \aCol{a}, reducing the complexity to \landO{\vert \mathcal{S}_\aCol{a} \vert \cdot \log \vert \mathcal{S}_\aCol{a} \vert + \rSize{\aTab{R}}} for ordered relations.

\subsection{Inclusion Dependencies}
\label{sec:validation:ind}
In general, we can validate a single IND candidate \IND{\aTab{R}.\aCol{a}}{\aTab{S}.\aCol{x}} by building a hash set of \aCol{x}'s values and checking if each value of \aCol{a} is contained in this set.
Multiple encoding characteristics and statistics can be exploited to accelerate validation.
First, we often observe that $\rSize{\aTab{R}} \gg \rSize{\aTab{S}}$ when \aTab{R} is a fact table and \aTab{S} is a dimension table.
Thus, the set of \aCol{x}'s values is relatively small, and many tuples in \aTab{R} reference the same key in \aTab{S}.
If \aCol{a} is dictionary-encoded, we do not have to probe each tuple for containment, but only the dictionary entries for each segment.

Second, we can use minimum and maximum values and continuity for further optimization.
For instance, the IND \IND{\aTab{R}.\aCol{a}}{\aTab{S}.\aCol{x}} cannot hold if $\min(\aCol{a}) < \min(\aCol{x})$ or $\max(\aCol{a}) > \max(\aCol{x})$, which can easily be derived from the segment statistics or the dictionaries.
Furthermore, the IND must hold if $\min(\aCol{a}) \geq \min(\aCol{x})$, $\max(\aCol{a}) \leq \max(\aCol{x})$, and $\forall v \in [\min(\aCol{x}),\, \max(\aCol{x})] : v \in \aCol{x}$, \ie \aCol{x} contains continuous values.
This rather straightforward reformulation allows us to drastically improve the validation performance for integer data types, \eg numeric keys:
if $\max(\aCol{x}) - \min(\aCol{x}) = \rCard{\aCol{x}} + 1$, \aCol{x} must contain all values in $[\min(\aCol{x}),\, \max(\aCol{x})]$.
For unique columns, $\rCard{\aCol{x}} = \rSize{\aCol{x}}$.
Thus, if we know that \aCol{x} is a UCC, we can check for continuousness and ensure that \aCol{x}'s minimal and maximal values match \aCol{a}.

The uniqueness property can either be given by an already validated UCC or derived by applying the same techniques as for UCC validation (set construction with index optimization for range-partitioned keys, see \cref{sec:validation:ucc}).
In the latter case, we also detect a valid UCC on \aCol{x}, which we store as well and do not need to validate again if requested.
We only fall back to probing \aCol{a}'s values to the hash set if \aCol{x} is not continuous.

The general validation strategy of building a set for \aTab{S}.\aCol{x}'s values and probing \aTab{R}.\aCol{a}'s values has a complexity in \landO{\rSize{\aTab{S}} + \rSize{\aTab{R}}}.
If \aCol{x} is continuous but unsorted, we can omit the probing step and reduce the complexity to \landO{\vert \mathcal{S}_{\aTab{S}.\aCol{x}} \vert \cdot \log \vert \mathcal{S}_{\aTab{S}.\aCol{x}} \vert + \rSize{\aTab{S}} + \vert \mathcal{S}_{\aTab{R}.\aCol{a}} \vert}.
For a range-partitioned and continuous integer key \aCol{x}, the complexity further decreases to \landO{\vert \mathcal{S}_{\aTab{S}.\aCol{x}} \vert \cdot \log  \vert \mathcal{S}_{\aTab{S}.\aCol{x}} \vert + \vert \mathcal{S}_{\aTab{R}.\aCol{a}} \vert} using the segment index or \landO{\vert \mathcal{S}_{\aTab{S}.\aCol{x}} \vert + \vert \mathcal{S}_{\aTab{R}.\aCol{a}} \vert} if we already validated that \aCol{x} is unique.
Identifying \aCol{a}'s minimum and maximum value in \landO{\vert \mathcal{S}_{\aTab{R}.\aCol{a}} \vert} is always required.
However, $\vert \mathcal{S}_{\aTab{R}.\aCol{a}} \vert  \ll \rSize{\aTab{R}}$ and $\vert \mathcal{S}_{\aTab{S}.\aCol{x}} \vert  \ll \rSize{\aTab{S}}$.
For foreign keys in fact and dimension tables, $\rSize{\aTab{S}} \ll \rSize{\aTab{R}}$.

\subsection{Ordering Dependency Candidates}
\label{sec:validation:candidate-order}
From the previous description of validation techniques, we observe characteristics that yield rules for a beneficial order to validate dependency candidates.
First, validating an IND candidate \IND{\aTab{R}.\aCol{a}}{\aTab{S}.\aCol{x}} always confirms a possible UCC \aTab{S}.\aCol{x}.
If this UCC is also a candidate, we can skip its validation later.
Second,  we can skip the validation of an FD candidate if any of the candidate columns is a UCC\@.

Furthermore, the rules generating dependency candidates can provide additional information.
The candidate rule for O-3 is an example of \emph{candidate dependence}.
\Cref{sec:background:rewrites} explains that we need an OD, an IND, and a UCC to apply this rewrite based on a range predicate.
If it is invalid, an OD candidate \OD{\aTab{S}.\aCol{x}}{\aTab{S}.\aCol{y}} can be rejected early using sampling.
An IND candidate \IND{\aTab{R}.\aCol{a}}{\aTab{S}.\aCol{x}} cannot be rejected before constructing \aTab{S}.\aCol{x}'s value set if \aTab{S}.\aCol{x} does not contain all values present in \aTab{R}.\aCol{a}.
Thus, the need to validate the IND candidate depends on the validation result of the OD candidate.
We track this dependence and only validate the IND if the OD has not been rejected before.
Combining all these observations, we obtain a clear candidate order by dependency type: we validate ODs first, INDs second, UCCs third, and FDs last.

\section{Building Blocks for System Integration}
\label{sec:building_blocks}
This section presents techniques required to leverage de\-pen\-den\-cy-based optimizations.
We describe how we propagate valid dependencies for each operator in the query plan during optimization.
This knowledge is crucial to return correct and complete query results.
Dependency-based optimizations rewrite joins into selections using the results of uncorrelated scalar subqueries (see join-to-predicate rewrite O-3 in \cref{sec:background:rewrites}), which require adjusted treatment in the query plan and enable further optimization during execution.
Thus, we propose dedicated subquery handling concepts to leverage dynamic pruning using subquery results at execution.

\subsection{Data Dependency Propagation}
\label{sec:propagation}
Evaluating the validity of dependencies for a specific logical operator is cumbersome, as operators can modify the required properties of a relation.
For instance, a UCC \aTab{R}.\aCol{a} might not be valid after an inner join \innerjoin{R}{S}, as each tuple in \aTab{R} can have multiple join partners, or an IND \IND{\aTab{R}.\aCol{a}}{\aTab{S}.\aCol{x}} can be invalid after a selection on \aTab{S}.
Thus, we adequately propagate dependencies in the query plan (C-2).
\citet{DBLP:journals/pvldb/LiuWSPLJYYLC23} gave anecdotal evidence that this propagation is not trivial by stating that implementing UCC propagation in Postgres lasted more than two years.\punctfootnote{See \url{https://commitfest.postgresql.org/35/2433/} \lastaccess.}

We achieve precise dependency information by consecutive dependency propagation and adaptation by each logical operator.
Starting from the declared or validated dependencies persisted for a relation (see \cref{sec:approach:overview}), each operator adds or removes dependencies. 
As query plans are subject to change for each optimization step,
operators do not persist dependencies but recursively compute them on the fly based on their input operators' dependencies.
The following paragraphs explain how we propagate the dependencies as displayed in \cref{tab:propagation}.

\begin{table}[tb]
    \centering
    \caption{Rules for dependency propagation. By default, operators forward input dependencies.}
    \label{tab:propagation}
    \resizebox{\linewidth}{!}{
    \begin{tabular}{@{}ccc@{}}
        \toprule
        \textbf{Operator} & \textbf{Input dependencies} & \textbf{Output dependencies} \\
        \midrule
        \multicolumn{3}{c}{\emph{Unique column combinations}} \\
            $\aggregation{R}{\aCol{a},\,\aCol{b},\,\aCol{sum(c)}}$ & --- & $\left\{{\aCol{a}, \aCol{b}}\right\}$ \\
            $\aggregation{R}{\aCol{avg(a)}}$ & --- & \aCol{avg(a)}  \\
            $\innerjoin[\aTab{R}.\aCol{a} \pEq \aTab{S}.\aCol{x}]{R}{S}$ & \aTab{R}.\aCol{b}, \aTab{S}.\aCol{x} & \aCol{b}, \FD{\aCol{x}}{\aTab{S} \setminus \aCol{x}} \\
            $\innerjoin[\aTab{R}.\aCol{a} \pEq \aTab{S}.\aCol{x}]{R}{S}$ &  \aTab{R}.\aCol{b}, \aTab{S}.\aCol{y} & \FD{\aCol{b}}{\aTab{R} \setminus \aCol{b}}, \FD{\aCol{y}}{\aTab{S} \setminus \aCol{y}} \\
            $\innerjoin[\aTab{R}.\aCol{a} \,\theta\, \aTab{S}.\aCol{x}]{R}{S}$ & \aTab{R}.\aCol{b}, \aTab{S}.\aCol{y} & \FD{\aCol{b}}{\aTab{R} \setminus \aCol{b}}, \FD{\aCol{y}}{\aTab{S} \setminus \aCol{y}} \\
            $\leftjoin[\aTab{R}.\aCol{a} \pEq \aTab{S}.\aCol{x}]{R}{S}$ & \aTab{R}.\aCol{b}, \aTab{S}.\aCol{x} & \FD{\aCol{b}}{\aTab{R} \setminus \aCol{b}}, \FD{\aCol{y}}{\aTab{S} \setminus \aCol{y}} \\
            $\aTab{R} \cup \aTab{S}$ & \aTab{R}.\aCol{a}, \aTab{S}.\aCol{x} & --- \\
        \midrule
        \multicolumn{3}{c}{\emph{Order dependencies}} \\
            $\innerjoin[\aTab{R}.\aCol{a} \pEq \aTab{S}.\aCol{x}]{R}{S}$ & --- & \OD{\aCol{a}}{\aCol{x}}, \OD{\aCol{x}}{\aCol{a}}  \\
            $\aTab{R} \cup \aTab{S}$ & \OD{\aTab{R}.\aCol{a}}{\aTab{R}.\aCol{b}}, \OD{\aTab{S}.\aCol{a}}{\aTab{S}.\aCol{b}} & --- \\
        \midrule
            \multicolumn{3}{c}{\emph{Inclusion dependencies}} \\
            $\selection{S}{\ensuremath{\aCol{y} \,\theta\, c }}$ & \IND{\aTab{R}.\aCol{a}}{\aTab{S}.\aCol{x}} & --- \\
            $\selection{S}{\aCol{x} \pIsNotNull}$ & \IND{\aTab{R}.\aCol{a}}{\aTab{S}.\aCol{x}} & \IND{\aTab{R}.\aCol{a}}{\aTab{S}.\aCol{x}} \\
            $\innerjoin{S}{T}$ & \IND{\aTab{R}.\aCol{a}}{\aTab{S}.\aCol{x}} & --- \\
            $\leftjoin{S}{T}$ & \IND{\aTab{R}.\aCol{a}}{\aTab{S}.\aCol{x}} & \IND{\aTab{R}.\aCol{a}}{\aTab{S}.\aCol{x}} \\
            $\innerjoin[\aTab{S}.\aCol{y} \pEq \aTab{T}.\aCol{u}]{S}{T}$ & \IND{\aTab{R}.\aCol{a}}{\aTab{S}.\aCol{x}}, \IND{\aTab{S}.\aCol{y}}{\aTab{T}.\aCol{u}} & \IND{\aTab{R}.\aCol{a}}{\aTab{S}.\aCol{x}} \\
         \bottomrule
    \end{tabular}
    }
\end{table}

\myparagraph{Unique column combinations}
We forward UCCs if all required columns are part of the operator's output and no function modifies the values.
A UCC \aCol{a} on \aTab{R} is invalidated (i)~by inner equi-joins \innerjoin{R}{S} where \aTab{S}'s join key is not unique, (ii)~by outer and theta-joins,  and (iii)~by unions.
However, new UCCs arise (i)~for grouping columns/distinct selections and (ii)~for ungrouped aggregates.

\myparagraph{Functional dependencies}
FDs can always be derived from existing UCCs and ODs.
Even after joins \innerjoin[\aTab{R}.\aCol{a} \pEq \aTab{S}.\aCol{x}]{R}{S} where \aCol{x} is not unique and after theta-joins, the UCC \aCol{b} yields the FD \FD{\aCol{b}}{\aTab{R} \setminus \aCol{b}}, which is forwarded.
These forwarded FDs remain unchanged as long as the involved attributes are part of the operator output. 

\myparagraph{Order dependencies}
ODs are invalidated by union operators or if their attributes are not part of the operator output. 
However, the join keys of an equi-join \innerjoin[\aTab{R}.\aCol{a} \pEq \aTab{S}.\aCol{x}]{R}{S} form two ODs \OD{\aCol{a}}{\aCol{x}} and \OD{\aCol{x}}{\aCol{a}}, as seen in \cref{tab:propagation}.
For such joins, existing ODs with the join key(s) on the left-hand side form transitive ODs with the other relation's join key(s).
As we derive FDs from ODs, this behavior reflects transitive FDs for the join keys.

\myparagraph{Inclusion dependencies}
For an IND \IND{\aTab{R}.\aCol{a}}{\aTab{S}.\aCol{x}}, it is not obvious whether it should be propagated starting from \aTab{R} or \aTab{S}.
Furthermore, INDs are also the most volatile dependency type: a single selection on \aTab{S} can invalidate the IND\@.
Thus, we persist them as a dependency on both relations and propagate them starting at \aTab{S}.
To prove whether a propagated IND holds, we must only check whether all foreign key columns are still present in the plan. 
We forward INDs if all columns are part of an operator's output, except for selections and filtering joins.
Selections only propagate an input IND with \aCol{x} as the referenced column for \selection{S}{x \pIsNotNull}.
In most cases, selections return an empty set of INDs and do not recurse further.

By incorporating dependency propagation in the query plan, we enable dependency-based query optimization out of the box and move further to making data dependencies first-class citizens of the database system.

\subsection{Subquery Handling}
\label{sec:subqueries}
Usually, database optimizers rewrite subqueries to (semi-)joins (\emph{subquery unnesting}) to avoid evaluating the subquery for each row~\cite{DBLP:journals/tods/Kim82,DBLP:conf/sigmod/HaasFLP89,DBLP:journals/pvldb/BellamkondaAWAZL09,DBLP:conf/sigmod/GanskiW87}.
However, executing predicates containing the result of an uncorrelated scalar subquery is more efficient, as we can execute the subquery once and use its result like a regular constant.
Thus, we must handle these predicates accordingly in the query plan during optimization (C-3), even though the exact predicate values are unknown until execution.
We identify two main challenges to employing these ideas: (i)~cardinality estimation and (ii)~partition pruning.
Our proposed solutions are  generally applicable and not limited to facilitating dependency-based optimization.

\myparagraph{Cardinality estimation}
\label{sec:subqueries:cardinality}
Cardinality estimation of predicates using subquery results is necessary to usefully place (semi-)joins rewritten to predicates (see \cref{fig:rewrites-subfig:o3i,fig:rewrites-subfig:o3ii}) in the query plan.
However, the results of scalar subqueries are unknown before execution.
While simply calculating the cardinality of equality predicates (\cref{fig:rewrites-subfig:o3i}) using the column's distinct value count is a well-known coarse estimate~\cite{DBLP:conf/sigmod/SelingerACLP79}, this approach does not work for range predicates with unknown lower and upper bounds (\cref{fig:rewrites-subfig:o3ii}).

Thus, we leverage the knowledge that the predicates generated by O-3 originated as (semi-)joins:
whenever the pattern of a subquery predicate matches the rewrite, we perform an estimation as for the original semi-join (\cref{fig:rewrites-subfig:o2}).
The optimizer can use its regular estimation techniques, which are often histogram-based~\cite{DBLP:journals/pvldb/WangQWWZ21}.
Without knowing the exact predicate values, the estimate for the semi-join is probably the closest approximation the cardinality estimator can provide.
In particular, it matches the estimation without applying O-3, leading to similar placement in the plan.
Different placements can effectively alter the join order, causing rather different, probably less beneficial query plans~\cite{DBLP:conf/vldb/ReddyH05}.

\myparagraph{Partition pruning using subquery results}
\label{sec:subqueries:pruning}
Commonly, horizontally partitioned databases prune partitions based on statistics, such as zone maps or range sets~\cite{DBLP:journals/pvldb/ZiauddinWKLPK17,DBLP:journals/pvldb/OrrKC19,DBLP:journals/pvldb/NicaSACHBG17}.
The operator accessing the data first skips partitions where no tuples can match selection predicates, reducing the amount of data being processed by all operators in the plan.
Clearly, we cannot determine pruning criteria during optimization when the predicate values are yet to be determined by query execution.
However, we keep track of the predicates that could enable pruning and shift from static partition pruning during optimization to \emph{dynamic} pruning using subquery results during execution.

We link predicates with scalar subquery results to the operators that first access the base relations by collecting the predicates from subsequent operators.
Imagine there are further operators in \cref{fig:rewrites-subfig:o3i} before the predicate on \aCol{s\_sold\_date}, \eg other selections or filtering semi-joins.
Then, the first operator on \aTab{sales} is enriched with additional predicates that enable subquery pruning.
When scheduling physical operators for execution, we add the operators that determine the predicate value as predecessors of the operator accessing the \aTab{sales} relation.
Thus, the subquery is executed first, and we can perform dynamic partition pruning with the evaluated subquery results when executing the operator loading the data.
We take care to guarantee that the resulting operator graphs are acyclic to avoid mutual waiting situations.
Cycles could happen due to subplan deduplication, where equivalent subplans are mapped to a single operator sequence.

Our proposed technique for dynamic partition pruning using subquery results is generally applicable to predicates using results of uncorrelated scalar subqueries.
Such predicates, \eg data-induced predicates (diPs)~\cite{DBLP:journals/pvldb/OrrKC19}, can be further optimized with this form of dynamic pruning.

\section{Evaluation}
\label{sec:evaluation}
In this section, we evaluate the impact of dependency-based optimizations and the efficiency of metadata-aware data dependency validation (C-4).
After briefly describing four standard benchmarks and characteristics of our experimental environment,
we study the impact of dependency-based optimizations on four different DBMSs through SQL rewrites and compare the performance to optimization integrated into a DBMS.
Then, we evaluate the performance impact per optimization technique and benchmark in the context of the additional dependency discovery overhead for Hyrise. 
Finally, we analyze the benefits of metadata-aware dependency validation algorithms and discuss the experimental results.

\subsection{Experimental Setup}
\label{sec:evaluation:setup}
We evaluate our approach using four standard benchmarks.
Besides the industry-standard TPC-H~\cite{TPCH} and TPC-DS~\cite{TPCDS} benchmarks (limited to 48~TPC-DS queries supported by Hyrise), we use the star schema benchmark (SSB)~\cite{DBLP:conf/tpctc/ONeilOCR09,SSB}  and the join order benchmark (JOB)~\cite{DBLP:journals/pvldb/LeisGMBK015}.
TPC-H, TPC-DS, and SSB allow controlling the amount of data using a scale factor (SF).
If not stated differently, this SF is 10.
JOB is based on the fixed real-world IMDB dataset, so it does not provide scaling.
We conducted the experiments on one non-uniform memory access (NUMA) region of an Ubuntu~24.04~LTS server with an Intel Xeon Platinum~8180 CPU (28~cores/56~threads) and \qty{378}{GiB} of local memory.
Our Hyrise plug-in was implemented in C++ and compiled using LLVM-17.
To ease interpretation, we use \emph{symmetric logarithmic axes}~\cite{journals/mscit/Webber13} in \cref{fig:performance,fig:validation_times}, which are linear close to 0 and logarithmic for larger values.

\subsection{Optimization for Different DBMSs}
\label{sec:evaluation:systems}
We show the potential of dependency-based QO for five analytical DBMSs:
\emph{DuckDB}~\cite{DBLP:conf/sigmod/RaasveldtM19} (1.1.3),
\emph{MonetDB}~\cite{DBLP:journals/debu/IdreosGNMMK12} (11.51.3), \emph{Umbra}~\cite{DBLP:conf/cidr/NeumannF20} (24.11),
\emph{Hyrise}~\cite{DBLP:conf/edbt/DreselerK0KUP19}, and
the commercial in-memory DBMS 
\emph{\hana}~\cite{DBLP:journals/debu/FarberMLGMRD12,DBLP:journals/sigmod/FarberCPBSL11,DBLP:conf/btw/May0L17}.
We conducted the experiments for \hana on a cloud instance with 128 vCores (Intel Xeon Platinum~8260) and \qty{1008}{GiB} of RAM.

We do not aim to compare system performance but the improvements achieved by dependency-based query optimization.
Thus, we report only the relative runtime improvement per system.
Similar to real-world applications~\cite{DBLP:conf/icde/SethiTSPXSYJHSB19,DBLP:journals/pvldb/RenenHPVDNLSKK24,DBLP:conf/sigmod/ArmenatzoglouBB22},
we consider a multi-threaded, high-load scenario with 32~concurrent clients.
Each client executes permutations of all TPC-H, TPC-DS, SSB,
and JOB queries
for two hours.
We measure the median runtime of all complete workload executions using four configurations:
first, no primary and foreign keys are specified as a baseline.
Second, we provide PKs and FKs. The systems can utilize index-based operators and own de\-pen\-den\-cy-based optimizations.
Third, we reformulate the SQL queries with optimizations O-1 (dependent group-by reduction) and O-3 (join-to-pre\-di\-cate rewrite).
We split subqueries generated by O-3 into separate statements and insert the subquery results as concrete predicate values to prevent the DBMSs from unnesting the subqueries to the original joins.
Finally, we combine the latter two configurations.
Because subquery results are unknown at optimization time, this configuration demonstrates the potential of the optimizations as an upper bound, where systems can apply the best predicate orders.

\begin{figure}[tb]
    \centering
    \includegraphics[width=\linewidth]{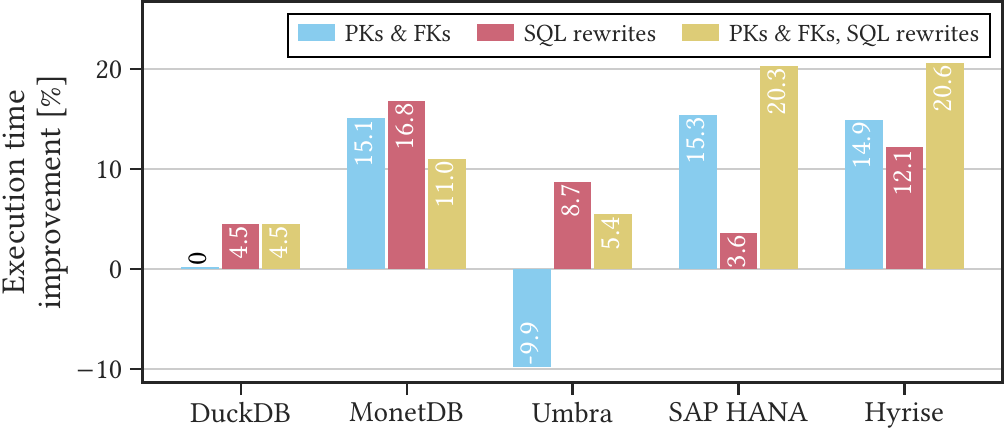}
    \caption{Execution time improvement of de\-pen\-den\-cy-based optimization by PKs/FKs and SQL queries over baselines without schema constraints (MT).}
    \label{fig:systems_comparison}
\end{figure}

\Cref{fig:systems_comparison} shows  the relative runtime improvements per configuration over the systems' baselines.
Only DuckDB and  Umbra show no considerable improvements when providing PKs and FKs.
We assume that Umbra's performance degradation is caused by the optimizer choosing index-based operators also when they are not beneficial.
Using only SQL rewrites benefits Umbra and MonetDB more than additionally specifying keys, likely for the same reason.
This behavior highlights that applying dependency-based optimizations is not trivial, and some systems (\eg DuckDB) do not implement them at all.
For \hana and Hyrise, the performance is best when adding SQL rewrites to schema constraints.
However, we can improve the runtime compared to schema constraints for all systems.
For \hana and Hyrise, we reach the best performance when combining schema-defined constraints and SQL rewrites.
Our integrated and automatic solution for Hyrise closely resembles this upper bound with an improvement of \perc{18.3}, illustrating that our system integration performs well.

\subsection{Performance Impact and Dependency Discovery Overhead}
\label{sec:evaluation:performance}
We compare the overhead of dependency discovery to the performance benefit of the three rewrites O-1 to O-3 (\cref{sec:background:rewrites}) for Hyrise.
We perform single-threaded (ST) and multi-threaded (MT) benchmark executions with one client.
The single-threaded setup allows us to assess the efficiency of optimized query plans without hiding latency by parallelism.
Multi-threaded experiments are limited to the NUMA region's 28~physical cores to ensure stable measurements.
We report the average latency of 100~repetitions within a time limit of \s{60} per query.
For the baseline execution (\emph{W/o deps.}), we do not provide any schema-defined primary and foreign key constraints.
We expect the rewrites to have a varying performance impact based on the benchmark characteristics.

\Cref{tab:performance} depicts the latency impact of the three optimizations and their combination for four benchmarks compared to the dependency discovery overhead (candidate generation and validation).
In general, the overhead of dependency discovery is (much) lower than the saved latency already for a single benchmark execution if there are valid candidates.
We achieve an average latency improvement of at least \perc{9} (\perc{6} MT) through all benchmarks when combining all optimization techniques, where the 
overhead is at least one order of magnitude smaller than the saved execution time.
Furthermore, we report the execution time with combined optimization techniques when knowing all schema-defined dependencies and the additional improvement enabled by further dependencies.

\begin{table*}[t]
    \centering\setlength{\tabcolsep}{3pt}
    \caption{Performance impact and overhead of dependency discovery for four benchmarks and three rewrite techniques O-1 to O-3 (see~\cref{sec:background:rewrites}), as well as all techniques combined, with schema-defined and additional dependencies. Overall single- (ST) and multi-threaded ({\color{gray}MT}) execution time in seconds [\unit{\second}] and relative latency change [\unit{\percent}] (one client). For dependency discovery, $\boldsymbol{\#}$ is the number of dependency candidates, $\boldsymbol{\checkmark}$ is the number of valid candidates, and \unit{\milli\second} is the total discovery time.}
    \label{tab:performance}
    \resizebox{\linewidth}{!}{
    \begin{tabular}{@{}rr@{~}r>{\color{gray}}r@{~}>{\color{gray}}rrrr|r@{~}r>{\color{gray}}r@{~}>{\color{gray}}rrrr|r@{~}r>{\color{gray}}r@{~}>{\color{gray}}rrrr|r@{~}r>{\color{gray}}r@{~}>{\color{gray}}rrrr@{}}
        \toprule
        
        & \multicolumn{7}{c}{TPC-H (22 queries)} & \multicolumn{7}{c}{TPC-DS (48 queries)} &  \multicolumn{7}{c}{SSB (13 queries)} &  \multicolumn{7}{c}{JOB (113 queries)} \\
        \cmidrule(lr){2-8}\cmidrule(lr){9-15}\cmidrule(lr){16-22}\cmidrule(lr){23-29}
        & \multicolumn{4}{c}{Execution} & \multicolumn{3}{c}{Discovery}
        & \multicolumn{4}{c}{Execution} & \multicolumn{3}{c}{Discovery} 
        & \multicolumn{4}{c}{Execution} & \multicolumn{3}{c}{Discovery}
        & \multicolumn{4}{c}{Execution} & \multicolumn{3}{c}{Discovery}  \\
        \cmidrule(lr){2-5}\cmidrule(lr){6-8}\cmidrule(lr){9-12}\cmidrule(lr){13-15}\cmidrule(lr){16-19}\cmidrule(lr){20-22}\cmidrule(lr){23-26}\cmidrule(lr){27-29}
        & \multicolumn{2}{c}{ST [\unit{\second} (\unit{\percent})]} & \multicolumn{2}{c}{\color{gray}MT [\unit{\second} (\unit{\percent})]} & \multicolumn{1}{c}{$\#$} & \multicolumn{1}{c}{\checkmark} & \multicolumn{1}{c|}{\unit{\milli\second}} 
        & \multicolumn{2}{c}{ST [\unit{\second} (\unit{\percent})]} & \multicolumn{2}{c}{\color{gray}MT [\unit{\second} (\unit{\percent})]} & \multicolumn{1}{c}{$\#$} & \multicolumn{1}{c}{\checkmark} & \multicolumn{1}{c|}{\unit{\milli\second}} 
        & \multicolumn{2}{c}{ST [\unit{\second} (\unit{\percent})]} & \multicolumn{2}{c}{\color{gray}MT [\unit{\second} (\unit{\percent})]} & \multicolumn{1}{c}{$\#$} & \multicolumn{1}{c}{\checkmark} & \multicolumn{1}{c|}{\unit{\milli\second}} 
        & \multicolumn{2}{c}{ST [\unit{\second} (\unit{\percent})]} & \multicolumn{2}{c}{\color{gray}MT [\unit{\second} (\unit{\percent})]} & \multicolumn{1}{c}{$\#$} & \multicolumn{1}{c}{\checkmark} & \multicolumn{1}{c}{\unit{\milli\second}} \\
        \midrule
        
        W/o deps.   & \num{37.5} &           &   \num{18.1} &              &    &    &    & \num{38.1} &             & \num{16.8}   &              &    &    &    &   \num{10.4} &              &    \num{4.0} &              &    &   &    &   \num{33.3} &              &   \num{20.6} &              &    &    &     \\
        O-1        & \num{-1.9} & (\num{-5}) &   \num{-0.6} & (\num{-3})   &  9 &  4 & <1 & \num{-1.0} & (\num{-3})  & \num{\pm0.0} & (\num{\pm0}) & 32 &  2 & <1 & \num{\pm0.0} & (\num{\pm0}) & \num{\pm0.0} & (\num{\pm0}) &  7 & 0 & <1 & \num{\pm0.0} & (\num{\pm0}) & \num{\pm0.0} & (\num{\pm0}) &  0 &  0 &   <1 \\
        O-2        & \num{-1.8} & (\num{-5}) &   \num{-0.3} & (\num{-1})   &  6 &  6 & <1 & \num{-7.0} & (\num{-18}) & \num{-1.9}   & (\num{-11})  & 15 & 11 & <1 &   \num{-1.2} & (\num{-11})  &   \num{-0.5} & (\num{-13})  &  4 & 4 & <1 &   \num{-4.2} & (\num{-13})  &   \num{-3.2} & (\num{-15})  & 10 & 10 & 260 \\
        O-3        & \num{-0.6} & (\num{-1}) & \num{\pm0.0} & (\num{\pm0}) & 22 &  9 & <1 & \num{-7.1} & (\num{-19}) & \num{-2.0}   & (\num{-12})  & 46 & 16 & 13 &   \num{-0.1} & (\num{-1})   &   \num{-0.2} & (\num{-5})   & 15 & 6 & 10 &   \num{-5.3} & (\num{-16})  &   \num{-3.2} & (\num{-15})  & 38 & 15 &  33 \\
        Combined   & \num{-3.6} & (\num{-9}) &   \num{-1.0} & (\num{-6})   & 31 & 13 & <1 & \num{-9.5} & (\num{-25}) & \num{-2.6}   & (\num{-15})  & 85 & 25 & 13 &   \num{-1.2} & (\num{-12})  &   \num{-0.7} & (\num{-17})  & 22 & 7 & 10 &   \num{-6.6} & (\num{-20})  &   \num{-4.5} & (\num{-22})  & 40 & 17 & 285 \\
        \midrule
        PKs \& FKs & \num{34.5} &            &  \num{17.0}  &              &    &    &    & \num{30.6} &             & \num{14.7}   &              &    &    &    &    \num{9.5} &              &    \num{3.5} &              &    &   &    &   \num{29.0} &              &  \num{17.4}  &              &    &    &     \\
        + UCCs, ODs& \num{-0.5} & (\num{-1}) & \num{\pm0.0} & (\num{\pm0}) & 24 &  3 & <1 & \num{-2.0} & (\num{-6})  & \num{-0.5}   & (\num{-3})   & 74 &  4 & 12 &   \num{-0.3} & (\num{-3})   &   \num{-0.1} &  (\num{-4})  & 18 & 2 & <1 &   \num{-2.3} & (\num{-8})   &   \num{-1.3} & (\num{-7})   & 30 &  7 &  21 \\
        \bottomrule
    \end{tabular}
    }
\end{table*}

The impact of O-2 (join-to-semi-join rewrite) and O-3 (join-to-predicate rewrite) is high for TPC-DS and JOB\@.
These benchmarks have snowflake schemas, which result in many joins of fact tables and dimension tables that can be rewritten.
Each join rewritten to a predicate by O3 can also be turned into a semi-join by O-2.
Thus, the impact of the two rewrites does not add up when all optimizations are applied.
O-1 (dependent group-by reduction) is most beneficial for TPC-H, where aggregates are also dominant~\cite{DBLP:journals/pvldb/DreselerBRU20}.

Exploiting dependencies beyond the schema (\emph{+ UCCs, ODs} in \cref{tab:performance}) yields further improvements for TPC-DS and JOB\@:
17~TPC-DS and 66~JOB queries improve with geometric mean speedups of \perc{35} and \perc{29}, respectively.
O-3 is the only optimization requiring more than schema-defined dependencies, and benchmarks where this optimization is beneficial profit most from discovering additional dependencies.
For instance, TPC-DS's Q37 has a latency improvement over \perc{90}.
The discovery time for all workloads is negligible when schema dependencies are known.

\begin{figure}[tb]
    \centering
    \begin{subfigure}{.475\linewidth}
        \centering
        \includegraphics[width=\linewidth]{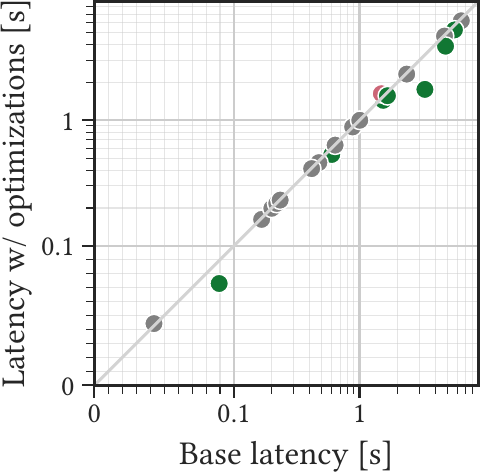}
        \caption{TPC-H (\perc{9})}
        \label{fig:performance-subfig:tpch}
    \end{subfigure}\hfill%
    \begin{subfigure}{.475\linewidth}
        \centering
        \includegraphics[width=\linewidth]{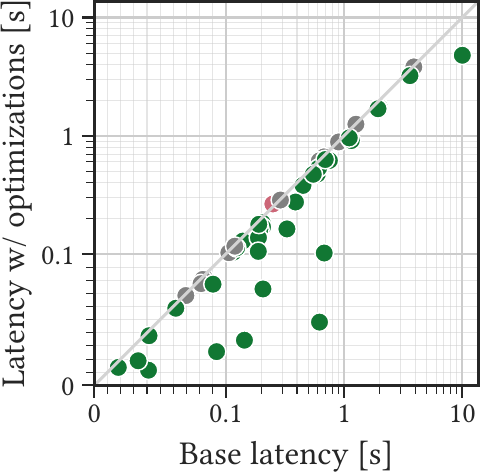}
        \caption{TPC-DS (\perc{25})}
        \label{fig:performance-subfig:tpcds}
    \end{subfigure}
    \\[2ex]
    \begin{subfigure}{.475\linewidth}
        \centering
        \includegraphics[width=\linewidth]{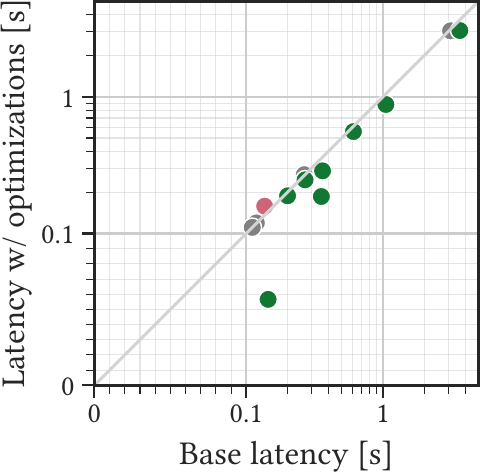}
        \caption{SSB (\perc{12})}
        \label{fig:performance-subfig:ssb}
    \end{subfigure}\hfill%
    \begin{subfigure}{.475\linewidth}
        \centering
        \includegraphics[width=\linewidth]{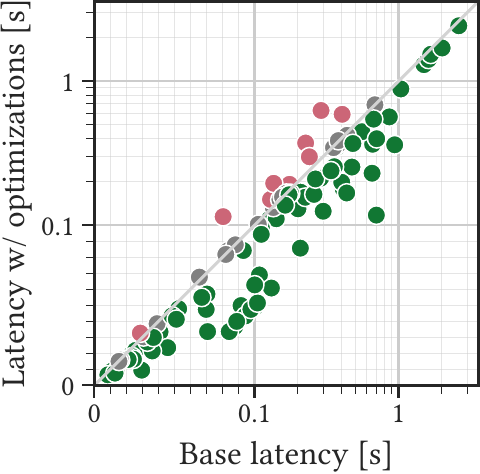}
        \caption{JOB (\perc{20})}
        \label{fig:performance-subfig:job}
    \end{subfigure}
    \caption{Latencies with and without de\-pen\-den\-cy-based optimizations per benchmark query (ST). Average relative latency improvement in parentheses. Queries that change at least by \textpm\perc{5} are colored (green/red).
    Note the bi-symmetric logarithmic axes (linear \s{<0.1}).}
    \label{fig:performance}
\end{figure}

\Cref{fig:performance} visualizes the ST performance impact on individual queries.
Each query is represented as a dot and placed with the baseline latency on the x-axis and the latency when applying the optimizations on the y-axis.
The optimizations improve query latency if the query is below the diagonal line.

For TPC-H, shown in \cref{fig:performance-subfig:tpch}, we observe that six out of 22 queries improve by at least \perc{5}, where Q10 benefits most from reducing seven group-by columns to one and decreases its latency by \perc{47}.
36 out of 48 TPC-DS queries improve by up to \perc{92} (\cref{fig:performance-subfig:tpcds}).
We observe high relative latency improvements when joins between fact tables and the \aTab{date} dimension are rewritten to range predicates and the physical order of tuples correlates to the date.
In this case, we can dynamically prune large parts of the fact table.
O-2 also achieves high absolute improvements and reduces the latency of Q95 by \ca{\s{4.9}} (\perc{51}).
However, Q1 degrades by \perc{8}: the optimizer does not place all semi-joins beneficially because of Hyrise's simple cost model.
We do not observe performance degradations from O-3 because our adapted subquery handling provides stable query plans compared to the original joins, \ie no join reordering. 
Eight out of 13 SSB queries improve by up to \perc{24}, whereas Q1.3 degrades (\cref{fig:performance-subfig:ssb}).
For JOB, 83 out of 113 queries improve up to \perc{83}, as shown in \cref{fig:performance-subfig:job}.
The UCC-based version of O-3 achieves high relative improvements.
Hyrise's cost-based semi-join pushdown is not always beneficial for this benchmark, as ten queries degrade.

\begin{figure}[tb]
    \centering
    \includegraphics[width=.475\linewidth]{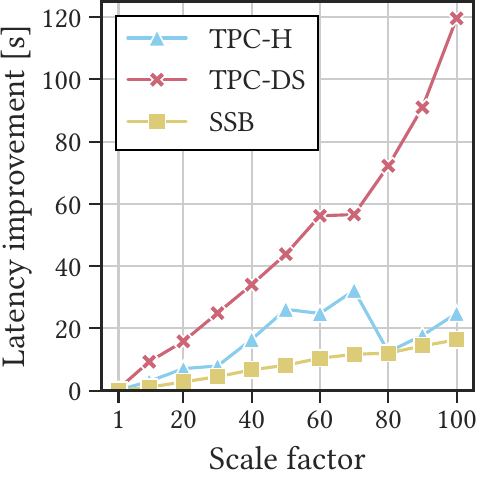}\hfill%
    \includegraphics[width=.475\linewidth]{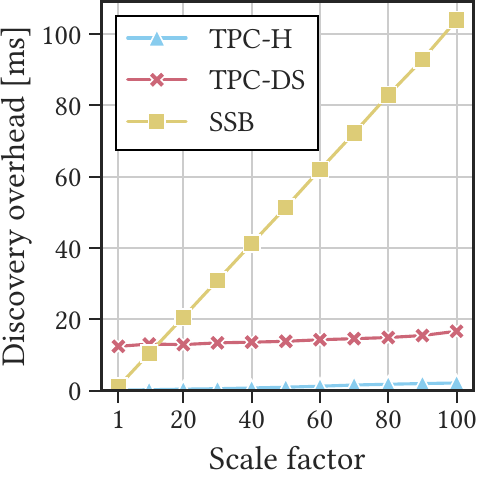}
    \caption{Latency improvement in seconds and discovery overhead in milliseconds for increasing scale factors (ST).}
    \label{fig:performance_tradeoff_sf}
\end{figure}

We also executed the workloads and dependency discovery using a scale factor (SF) of 1 and SFs from 20 to 100 for TPC-H, TPC-DS, and SSB.
The ST latency improvement compared to the dependency discovery overhead is shown in \cref{fig:performance_tradeoff_sf}.
For all scale factors, the dependency discovery overhead is orders of magnitude smaller than the latency improvement and does not exceed \ms{104} for SSB, \ms{3} for TPC-H, and \ms{17} for TPC-DS (all SF 100).
While this implies linear scaling for SSB and TPC-H, the discovery scales sub-linearly for TPC-DS\@:
all tables with candidates grow linearly with the scale factor for TPC-H and SSB, whereas some dimension tables of TPC-DS grow slower than linearly.
With dynamic subquery pruning (\cref{sec:subqueries:pruning}), 21 JOB queries improve latency by to \perc{31} and achieve a \perc{13} geometric mean speedup.

The latency improvement, \ie the saved execution time, has the highest growth rate for TPC-DS and the lowest growth rate for SSB\@.
For TPC-H, we observe a reduced latency improvement from SF~80 on, caused by previously reported disadvantageous placement decisions, most severely in Q21.
Here, rewritten semi-joins with large build sides are pushed below selections.

Our experiments demonstrate that dependency-based optimization techniques improve a database's performance.
Datasets with normalized schemas benefit more than datasets with fewer dimension tables because of the higher potential for join rewrites.
While exploiting schema-provided dependencies already has an impact, discovering and using additional dependencies can turn joins into selections, further improving performance.
Integration of subqueries used for these selections yields stable query plans compared to the baseline.
Discovering additional dependencies is amortized after a single benchmark execution.

\subsection{Metadata-aware Dependency Validation}
\label{sec:evaluation:validation}
We investigate the efficiency of the metadata-aware data dependency validation algorithms presented in \cref{sec:validation}
by reporting the validation times of all candidates generated for the benchmarks.

\begin{figure}[tb]
    \centering
    \includegraphics[height=0.475\linewidth]{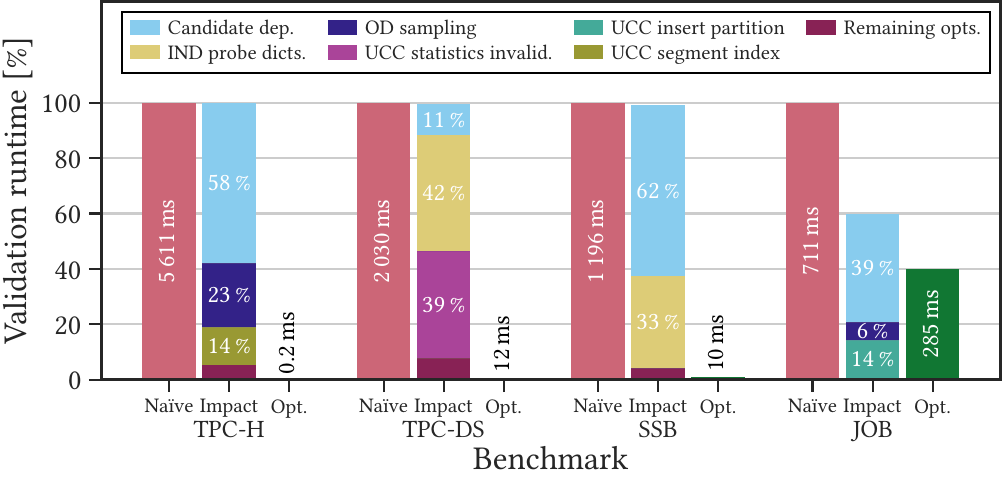}
    \caption{Dependency validation runtime improvement of metadata-aware, 
    optimized strategies based on impact of individual optimizations relative to the na\"ive baseline.
    }
    \label{fig:validation-improvement}
\end{figure}

\myparagraph{Performance impact of tailored validation}
We start with an ablation study to evaluate the impact of optimizations for dependency validation described in \cref{sec:validation}.
Instead, we use the fallback validation strategies as a baseline:
first,  we always construct a pre-al\-lo\-cated hash set to validate UCC candidates.
Second, we always build a hash set for the referenced column and probe all fields of the foreign key column for INDs.
Third, we always sort by the entire column for ODs.
Ultimately, we do not track candidate dependence and validate all INDs regardless of the validity of OD candidates.
In an ablation study, we activate all optimizations described in \cref{sec:validation} one by one and measure their impact (average of 100 executions).
\Cref{fig:validation-improvement} shows the overall validation times per benchmark using na\"ive and optimized validation techniques.
We observe a speedup of \num{2.5} for JOB and improvements of at least two orders of magnitude for the other benchmarks.

\Cref{fig:validation-improvement} also presents the impact of individual optimizations.
The contribution of optimizations varies between datasets, highlighting that their combination is necessary.
However, optimizations that are independent of data layout and encoding (candidate dependence, sampling for ODs, statistics-based invalidation for UCCs) decrease the validation time noticeably.
Optimizations exploiting dictionary encoding (probing only dictionaries for INDs, inserting entire segment to hash set for UCCs) or partitioning (segment index for UCCs) further improve validation performance.
The benefit of tailored validation techniques is 
most noticeably for TPC-H.
Here, our system generates IND candidates for the \aTab{lineitem} table and OD and UCC candidates on the \aTab{orders} table,
which are costly to validate by sorting an entire column or adding it to a hash set due to the relation sizes. 
We observe similar behavior for TPC-DS and SSB\@.
The validation improves less for JOB, as we often fall back to set-based validation techniques.

\begin{figure}[tb]
    \centering
    \begin{subfigure}{.475\linewidth}
        \centering
        \includegraphics[width=\linewidth]{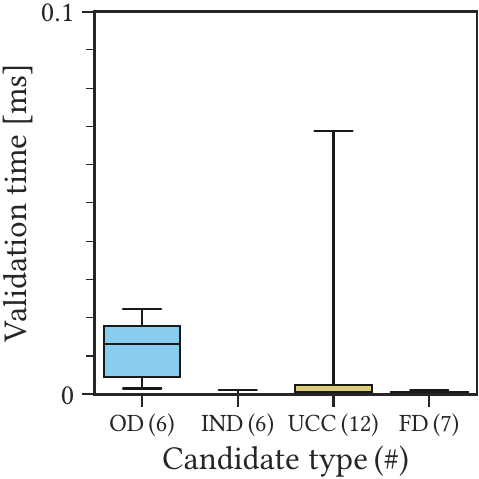}
        \caption{TPC-H (31)}
        \label{fig:validation_times-subfig:tpch}
    \end{subfigure}\hfill%
    \begin{subfigure}{.475\linewidth}
        \centering
        \includegraphics[width=\linewidth]{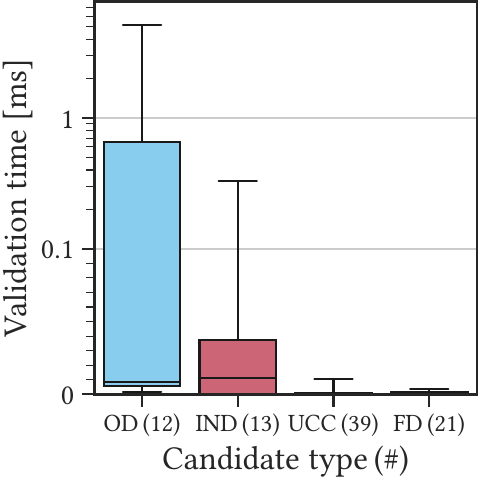}
        \caption{TPC-DS (85)}
        \label{fig:validation_times-subfig:tpcds}
    \end{subfigure}
    \\[2ex]
    \begin{subfigure}{.475\linewidth}
        \centering
        \includegraphics[width=\linewidth]{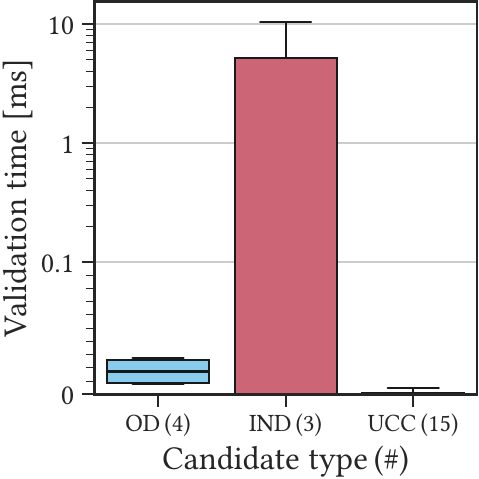}
        \caption{SSB (22)}
        \label{fig:validation_times-subfig:ssb}
    \end{subfigure}\hfill%
    \begin{subfigure}{.475\linewidth}
        \centering
        \includegraphics[width=\linewidth]{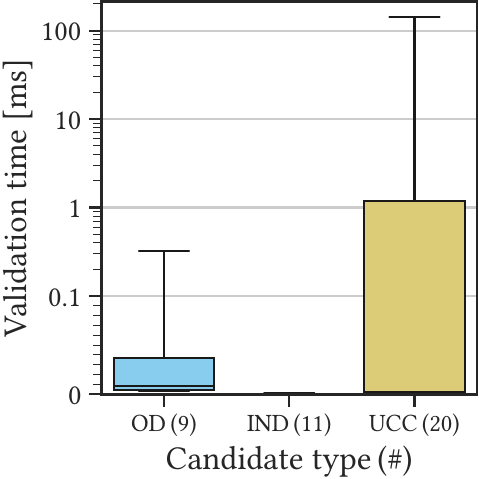}
        \caption{JOB (40)}
        \label{fig:validation_times-subfig:job}
    \end{subfigure}
    \caption{Average candidate validation runtimes for four benchmarks (number of candidates in parentheses). Note the symmetric logarithmic y-axis (linear \ms{<0.1}). Whiskers cover the entire value range.}
    \label{fig:validation_times}
\end{figure}

\myparagraph{Detailed validation performance per benchmark}
\Cref{fig:validation_times} shows the average validation time per candidate type and benchmark (average of \qty{1000}{executions} for stable results).
For the TPC-H benchmark, depicted in \cref{fig:validation_times-subfig:tpch}, the system generates 31~dependency candidates.
All candidates are validated in less than \mus{100}.
Five of six OD candidates are rejected after the sampling phase, and only \OD{\aTab{region}.\aCol{r\_regionkey}}{\aTab{region}.\aCol{r\_name}} is valid.
We skip the validation for five IND candidates as they depend on invalid ODs (see \cref{sec:validation:candidate-order}), although they represent foreign key relationships.
One of these skipped candidates is \IND{\aTab{lineitem}.\aCol{l\_orderkey}}{\aTab{orders}.\aCol{o\_orderkey}}.
This candidate would dominate the entire validation, with a validation time of \ca{\s{1}}.
The \aCol{o\_orderkey} column is not continuous, as only \perc{25} of the possible key range is populated~\cite[p.~86]{TPCH}.
Thus, we have to fall back to hash set construction and probing and cannot use optimizations using metadata.
However, we validate the remaining IND \IND{\aTab{nation}.\aCol{n\_regionkey}}{\aTab{region}.\aCol{r\_regionkey}} within microseconds.

We reject four of twelve UCC candidates in microseconds using metadata and skip another one on a primary key that was already confirmed as a byproduct of IND validation.
The remaining seven UCC candidates are range-partitioned primary keys that can be confirmed using metadata and the \aCol{name} columns of \aTab{nation} and \aTab{region}, which consist of a single partition.
The UCC on the \aTab{order} table's PK consumes the most time with \ca{\mus{70}}, as it has 229~segments that we index.
Three FD candidates are skipped due to containing a UCC, and the remaining four candidates are rejected by metadata.

\Cref{fig:validation_times-subfig:tpcds} shows the validation times of 85~candidates generated for the TPC-DS benchmark.
Eight out of twelve OD candidates are rejected by sampling. 
Each of the four remaining OD candidates has the \aTab{date\_dim}'s sequential key as the ordering column, which orders \aCol{d\_date}, the sequential month and quarter representation, and \aCol{d\_year}.
The two candidates involving string columns are confirmed in \ca{\ms{5}}.
All IND candidates represent valid foreign key relationships, but we skip the validation of five candidates that depend on invalid ODs.
We confirm the remaining eight candidates in under \ms{1} each by exploiting the sorted primary keys.
The validation time scales with the number of column segments, where the validation of \IND{\aTab{inventory}.\aCol{inv\_date\_sk}}{\aTab{date\_dim}.\aCol{d\_date\_sk}} takes \ca{\mus{300}} due to traversing \aCol{inv\_date\_sk}'s \num{2032} segments.
Sorted primary keys also enable the confirmation of eleven of the 39~UCC candidates within a few microseconds.
The IND validation already confirmed one UCC candidate, and the remaining 27~candidates are rejected by metadata immediately.
One FD candidate is confirmed by metadata in a few microseconds, and all 20~remaining candidates are immediately rejected by metadata.

The validation times for SSB, displayed in \cref{fig:validation_times-subfig:ssb}, show a larger variance.
Two OD candidates are rejected by sampling, and two ODs on the \aTab{date} table are confirmed in \ca{\mus{25}}.
The two invalid OD candidates lead to skipping two IND candidates, but the remaining candidate \IND{\aTab{lineorder}.\aCol{lo\_orderdate}}{\aTab{date}.\aCol{d\_datekey}} falls back to the hash set-based check because \aCol{d\_datekey} is not continuous~\cite[p.~4]{SSB}.
Thus, confirming this IND takes \ca{\ms{10}}.
We confirm three of 15 UCC candidates candidates and reject eleven by metadata within a few microseconds.
The UCC \aTab{date}.\aCol{d\_datekey} was already confirmed by IND validation.
No FD candidates were generated as no SSB query groups by multiple columns of the same table.

The validation of JOB's candidates takes the most time, as seen in \cref{fig:validation_times-subfig:job}.
All nine OD candidates are rejected by sampling in less than \ms{1}, allowing to skip all eleven IND candidates.
Three of the 20 UCC candidates are rejected by metadata, and twelve candidates are confirmed by metadata in microseconds.
The remaining five candidates are valid UCCs but are not range-partitioned.
Thus, we must use hash set construction for these candidates, taking up to \ms{105} for \aTab{char\_name}.\aCol{id} and \ms{142} for \aTab{name}.\aCol{id}.
The JOB queries have no group-by statements; thus, there are no FD candidates.

Our experiments confirm that metadata-aware validation is efficient for rejecting and confirming dependency candidates.
We observe the longest validation times for valid candidates that fall back to default validation, \ie complete sorting for ODs or hash set construction for UCCs and INDs.
Metadata-aware validation scales well with relation size and number of partitions.
Ordering the candidates by type allows skipping candidates known to be valid, and exploiting candidate dependence for candidates generated for O-3 reduces the validation overhead.

\subsection{Discussion}

Our experiments highlight that dependency-based query optimization techniques can improve database  performance.
Some degradation of individual queries arises from the interplay with other optimizer rules, which can be further tuned.

\myparagraph{Dependencies hold on dimension tables}
Most data dependencies required for the selected query rewrites are genuine dependencies on dimension tables, which are unlikely to be rendered invalid as data changes or are added.
However, the main limitation of our current solution is that dependency validation must be re-iterated as part of an ETL process whenever data changes. We discuss a corresponding extension as future work.

\myparagraph{Dimension table modeling uncovers potential}
If a dimension table's join key orders columns with selections, joins with the fact table can be reformulated to a scan of the fact table for both point and range predicates on the dimension.
Thus, the disadvantage of performing many costly joins when using a snowflake schema compared to a flat table~\cite{DBLP:conf/bigdataconf/BeiPATDBL19,DBLP:journals/is/LeveneL03} can be reduced.
In case the dimension table's key is also sequential and range-partitions the table, metadata-aware dependency validation can work solely on metadata, validating UCCs and INDs immediately without traversing entire columns.

\myparagraph{Schema normalization facilitates optimizations}
The schema design influences the potential of dependency-based optimizations and the efficiency of dependency discovery.
Normalized snowflake schemas with fact and dimension tables lead to more joins that can be reformulated, especially in combination with valid dependencies additional to PKs and FKs~(O-3).
We observe high relative improvements for short-running queries with few rewritten joins.
Long-running queries benefit if we rewrite many joins or a single dominating join (\eg TPC-DS's Q95).

\section{Conclusion and Future Work}
\label{sec:conclusion}
We evaluated three query optimizations using data dependencies that substantially reduce workload latencies, and presented methods for the discovery of relevant data dependencies within milliseconds.
We developed metadata-aware validation algorithms to confirm or reject data dependency candidates purely based on database statistics and described how to adapt a DBMS for thorough integration of dependency-based optimizations.
The methods to handle scalar subqueries in cardinality estimation and pruning are not limited to dependency-based query optimization.
Our experiments confirm the benefit of dependency-based query optimization four benchmarks and four DBMSs.
Compared to using only known primary/foreign keys, 17~TPC-DS and 66~JOB queries improve with \perc{35} and \perc{29} geometric mean speedups, respectively.

Our approach allows for perceiving data dependencies as pure metadata.
By eliminating the need to specify constraints only for the purpose of query optimization and providing automatic, workload-driven dependency discovery, we combine the concepts of autonomous databases~\cite{DBLP:conf/cidr/PavloAALLMMMPQS17} and data profiling.

We considered dependency discovery as part of ETL processes. 
Thus, a promising next step for future work is handling also frequently changing datasets with an online approach.

\balance

\bibliographystyle{ACM-Reference-Format}
\bibliography{sample}

\end{document}

%% file: plugin_architecture.pdf_tex
\begingroup%
  \makeatletter%
  \providecommand\color[2][]{%
    \errmessage{(Inkscape) Color is used for the text in Inkscape, but the package 'color.sty' is not loaded}%
    \renewcommand\color[2][]{}%
  }%
  \providecommand\transparent[1]{%
    \errmessage{(Inkscape) Transparency is used (non-zero) for the text in Inkscape, but the package 'transparent.sty' is not loaded}%
    \renewcommand\transparent[1]{}%
  }%
  \providecommand\rotatebox[2]{#2}%
  \newcommand*\fsize{\dimexpr\f@size pt\relax}%
  \newcommand*\lineheight[1]{\fontsize{\fsize}{#1\fsize}\selectfont}%
  \ifx\svgwidth\undefined%
    \setlength{\unitlength}{240.26400948bp}%
    \ifx\svgscale\undefined%
      \relax%
    \else%
      \setlength{\unitlength}{\unitlength * \real{\svgscale}}%
    \fi%
  \else%
    \setlength{\unitlength}{\svgwidth}%
  \fi%
  \global\let\svgwidth\undefined%
  \global\let\svgscale\undefined%
  \makeatother%
  \begin{picture}(1,0.5993407)%
    \lineheight{1}%
    \setlength\tabcolsep{0pt}%
    \put(0,0){\includegraphics[width=\unitlength,page=1]{plugin_architecture.pdf}}%
    \put(0.7480145,0.11718426){\color[rgb]{0,0,0}\makebox(0,0)[t]{\lineheight{1.25}\smash{\begin{tabular}[t]{c}\circled{7}\end{tabular}}}}%
    \put(0.10990099,0.4928793){\color[rgb]{0,0,0}\makebox(0,0)[t]{\lineheight{1.25}\smash{\begin{tabular}[t]{c}\footnotesize\textsf{SQL Query}\end{tabular}}}}%
    \put(0.24641606,0.50510247){\color[rgb]{0,0,0}\makebox(0,0)[lt]{\lineheight{1.25}\smash{\begin{tabular}[t]{l}\circled{1}\end{tabular}}}}%
    \put(0.23004291,0.56819022){\color[rgb]{0,0,0}\makebox(0,0)[lt]{\lineheight{1.25}\smash{\begin{tabular}[t]{l}\footnotesize\textsf{Database System}\end{tabular}}}}%
    \put(0.37692126,0.48679552){\color[rgb]{0,0,0}\makebox(0,0)[t]{\lineheight{.5}\smash{\begin{tabular}[t]{c}\footnotesize\textsf{Translation,}\\\footnotesize\textsf{Optimization}\end{tabular}}}}%
    \put(0.62374351,0.47742965){\color[rgb]{0,0,0}\makebox(0,0)[t]{\lineheight{1.25}\smash{\begin{tabular}[t]{c}\footnotesize\textsf{Query Plan}\end{tabular}}}}%
    \put(0.87056607,0.47734837){\color[rgb]{0,0,0}\makebox(0,0)[t]{\lineheight{1.25}\smash{\begin{tabular}[t]{c}\footnotesize\textsf{Execution}\end{tabular}}}}%
    \put(0.28148685,0.41177813){\color[rgb]{0,0,0}\makebox(0,0)[lt]{\lineheight{1.25}\smash{\begin{tabular}[t]{l}\circled{2}\end{tabular}}}}%
    \put(0.42400814,0.41177813){\color[rgb]{0,0,0}\makebox(0,0)[lt]{\lineheight{1.25}\smash{\begin{tabular}[t]{l}\circled{3}\end{tabular}}}}%
    \put(0.10653741,0.33157149){\color[rgb]{0,0,0}\makebox(0,0)[lt]{\lineheight{1.25}\smash{\begin{tabular}[t]{l}\circled{9}\end{tabular}}}}%
    \put(0.35350978,0.30776324){\color[rgb]{0,0,0}\makebox(0,0)[t]{\lineheight{1.25}\smash{\begin{tabular}[t]{c}\footnotesize\textsf{Dependency}\end{tabular}}}}%
    \put(0.87056607,0.36059456){\color[rgb]{0,0,0}\makebox(0,0)[t]{\lineheight{1.25}\smash{\begin{tabular}[t]{c}\footnotesize\textsf{Plan Cache}\end{tabular}}}}%
    \put(0.78452213,0.25182917){\color[rgb]{0,0,0}\makebox(0,0)[lt]{\lineheight{1.25}\smash{\begin{tabular}[t]{l}\circled{4}\end{tabular}}}}%
    \put(0.91357648,0.25182917){\color[rgb]{0,0,0}\makebox(0,0)[lt]{\lineheight{1.25}\smash{\begin{tabular}[t]{l}\circled{10}\end{tabular}}}}%
    \put(0.88785089,0.10330442){\color[rgb]{0,0,0}\makebox(0,0)[t]{\lineheight{.5}\smash{\begin{tabular}[t]{c}\footnotesize\textsf{Candidate}\\\footnotesize\textsf{Rule}\end{tabular}}}}%
    \put(0.57866405,0.14032759){\color[rgb]{0,0,0}\makebox(0,0)[t]{\lineheight{1.25}\smash{\begin{tabular}[t]{c}\circled{5}\end{tabular}}}}%
    \put(0.36752916,0.14032759){\color[rgb]{0,0,0}\makebox(0,0)[t]{\lineheight{1.25}\smash{\begin{tabular}[t]{c}\circled{6}\end{tabular}}}}%
    \put(0.62378716,0.0986224){\color[rgb]{0,0,0}\makebox(0,0)[t]{\lineheight{.5}\smash{\begin{tabular}[t]{c}\footnotesize\textsf{Candidate}\\\footnotesize\textsf{Generation}\end{tabular}}}}%
    \put(0.37533261,0.08145506){\color[rgb]{0,0,0}\makebox(0,0)[t]{\lineheight{1.25}\smash{\begin{tabular}[t]{c}\footnotesize\textsf{Candidate}\end{tabular}}}}%
    \put(0.02128193,0.02572443){\color[rgb]{0,0,0}\makebox(0,0)[lt]{\lineheight{1.25}\smash{\begin{tabular}[t]{l}\footnotesize\textsf{Dependency Discovery}\end{tabular}}}}%
    \put(0.10977276,0.09870355){\color[rgb]{0,0,0}\makebox(0,0)[t]{\lineheight{.5}\smash{\begin{tabular}[t]{c}\footnotesize\textsf{Dependency}\\\footnotesize\textsf{Validation}\end{tabular}}}}%
    \put(0.10878525,0.14032759){\color[rgb]{0,0,0}\makebox(0,0)[t]{\lineheight{1.25}\smash{\begin{tabular}[t]{c}\circled{8}\end{tabular}}}}%
    \put(0.65422965,0.26242017){\color[rgb]{0,0,0}\makebox(0,0)[t]{\lineheight{1.25}\smash{\begin{tabular}[t]{c}\footnotesize\textsf{Data}\end{tabular}}}}%
  \end{picture}%
\endgroup%

%% file: ucc_validation.pdf_tex
\begingroup%
  \makeatletter%
  \providecommand\color[2][]{%
    \errmessage{(Inkscape) Color is used for the text in Inkscape, but the package 'color.sty' is not loaded}%
    \renewcommand\color[2][]{}%
  }%
  \providecommand\transparent[1]{%
    \errmessage{(Inkscape) Transparency is used (non-zero) for the text in Inkscape, but the package 'transparent.sty' is not loaded}%
    \renewcommand\transparent[1]{}%
  }%
  \providecommand\rotatebox[2]{#2}%
  \newcommand*\fsize{\dimexpr\f@size pt\relax}%
  \newcommand*\lineheight[1]{\fontsize{\fsize}{#1\fsize}\selectfont}%
  \ifx\svgwidth\undefined%
    \setlength{\unitlength}{240.26399231bp}%
    \ifx\svgscale\undefined%
      \relax%
    \else%
      \setlength{\unitlength}{\unitlength * \real{\svgscale}}%
    \fi%
  \else%
    \setlength{\unitlength}{\svgwidth}%
  \fi%
  \global\let\svgwidth\undefined%
  \global\let\svgscale\undefined%
  \makeatother%
  \begin{picture}(1,0.31215664)%
    \lineheight{1}%
    \setlength\tabcolsep{0pt}%
    \put(0,0){\includegraphics[width=\unitlength,page=1]{ucc_validation.pdf}}%
    \put(0.51477842,0.19906908){\color[rgb]{0,0,0}\makebox(0,0)[rt]{\lineheight{1.25}\smash{\begin{tabular}[t]{r}\footnotesize$[1, 7]$\end{tabular}}}}%
    \put(0.51477843,0.02384306){\color[rgb]{0,0,0}\makebox(0,0)[rt]{\lineheight{1.25}\smash{\begin{tabular}[t]{r}\footnotesize$[12, 17]$\end{tabular}}}}%
    \put(0.61741658,0.0690363){\color[rgb]{0,0,0}\makebox(0,0)[t]{\lineheight{1.25}\smash{\begin{tabular}[t]{c}\vdots\end{tabular}}}}%
    \put(0.61777833,0.26666781){\color[rgb]{0,0,0}\makebox(0,0)[t]{\lineheight{1.25}\smash{\begin{tabular}[t]{c}\footnotesize\textsf{Segment}\end{tabular}}}}%
    \put(0.7431589,0.26662309){\color[rgb]{0,0,0}\makebox(0,0)[lt]{\lineheight{1.25}\smash{\begin{tabular}[t]{l}\footnotesize\textsf{Distinct values}\end{tabular}}}}%
    \put(0.51477843,0.26650116){\color[rgb]{0,0,0}\makebox(0,0)[rt]{\lineheight{1.25}\smash{\begin{tabular}[t]{r}\footnotesize$[\textsf{min, max}]$\end{tabular}}}}%
    \put(0.71793859,0.19910972){\color[rgb]{0,0,0}\makebox(0,0)[lt]{\lineheight{1.25}\smash{\begin{tabular}[t]{l}\footnotesize{$7$}\end{tabular}}}}%
    \put(0.71793859,0.13167764){\color[rgb]{0,0,0}\makebox(0,0)[lt]{\lineheight{1.25}\smash{\begin{tabular}[t]{l}\footnotesize{$7$}\end{tabular}}}}%
    \put(0.71793859,0.02388372){\color[rgb]{0,0,0}\makebox(0,0)[lt]{\lineheight{1.25}\smash{\begin{tabular}[t]{l}\footnotesize{$6$}\end{tabular}}}}%
    \put(0.51477843,0.131637){\color[rgb]{0,0,0}\makebox(0,0)[rt]{\lineheight{1.25}\smash{\begin{tabular}[t]{r}\footnotesize$[8, 14]$\end{tabular}}}}%
    \put(0.61777833,0.19923573){\color[rgb]{0,0,0}\makebox(0,0)[t]{\lineheight{1.25}\smash{\begin{tabular}[t]{c}\footnotesize\textsf{Segment 1}\end{tabular}}}}%
    \put(0.61777833,0.13180365){\color[rgb]{0,0,0}\makebox(0,0)[t]{\lineheight{1.25}\smash{\begin{tabular}[t]{c}\footnotesize\textsf{Segment 2}\end{tabular}}}}%
    \put(0.61777833,0.02400971){\color[rgb]{0,0,0}\makebox(0,0)[t]{\lineheight{1.25}\smash{\begin{tabular}[t]{c}\footnotesize\textsf{Segment 17}\end{tabular}}}}%
    \put(0.31995069,0.13967106){\color[rgb]{0,0,0}\makebox(0,0)[rt]{\lineheight{.5}\smash{\begin{tabular}[t]{r}\footnotesize\textsf{Tree-}\\\footnotesize\textsf{based}\\\footnotesize\textsf{index}\end{tabular}}}}%
  \end{picture}%
\endgroup%